\documentclass[letterpaper,twocolumn]{scrartcl}

\usepackage[utf8]{inputenc}
\usepackage[T1]{fontenc}

\usepackage{showyourwork}

\makeatletter
\@ifundefined{variable}{
}{}
\makeatother

\usepackage{aas_macros}
\usepackage{fontawesome5}

\usepackage[square,sort,comma,numbers]{natbib}

\usepackage{savesym}

\usepackage{graphicx}
\usepackage{bm}

\usepackage{soul}

\usepackage{pdflscape}
\usepackage{tabularx}

\usepackage{xcolor}
\usepackage{multirow}
\usepackage[ttdefault=true]{AnonymousPro}

\usepackage[frozencache]{minted}
\setminted[]{
  frame=lines,
  framesep=2mm,
  bgcolor=bg,
  baselinestretch=1.2,
  fontsize=\footnotesize,
  breaklines,
  breakafter=d,
  breakbefore=a,
}
\definecolor{bg}{rgb}{0.97, 0.97, 0.97}

\usepackage{ragged2e} \definecolor{commentsColor}{rgb}{0.497495, 0.497587, 0.497464}
\definecolor{keywordsColor}{rgb}{0.000000, 0.000000, 0.635294}
\definecolor{stringColor}{rgb}{0.558215, 0.000000, 0.135316}
\definecolor{backgroundColor}{rgb}{0.9, 0.9, 0.9}

\usepackage[most]{tcolorbox}

\usepackage{amssymb}

\usepackage{amsmath}

\usepackage{physics}

\usepackage{booktabs}

\usepackage{cancel}\usepackage{enumerate}\usepackage{calligra} \usepackage{mathtools} \usepackage{pgf}

\hypersetup{breaklinks,linktoc=all,colorlinks,urlcolor=blue,citecolor=blue,linkcolor=blue}

\usepackage{comment}
\usepackage{pgfpages}

\usepackage{savesym}

\usepackage{nicefrac}

\savesymbol{tablenum}
\usepackage{siunitx}
\restoresymbol{SIX}{tablenum}

\usepackage{relsize}

\definecolor{codegray}{gray}{0.885}
\renewcommand\t\text

\usepackage{tikz}
\savesymbol{plate}
\usetikzlibrary{bayesnet}

\usepackage[linesnumbered,lined,ruled,commentsnumbered]{algorithm2e}
\usepackage{xspace}

\usepackage{authblk}

\usepackage{fancyhdr}

\usepackage[sharp]{easylist}

\newcommand\pysr{\textsc{PySR}\xspace}
\newcommand\pysrjlabs{\mbox{\textsc{SymbolicRegression.jl}}\xspace}
\newcommand\pysrjl{\textsc{SymbolicRegression.jl}\xspace}
\newcommand\eureqa{\textsc{Eureqa}\xspace}
\newcommand\sr{SR\xspace}

\usepackage{cleveref} 
\begin{document}
\newcommand\myhref[2]{\textsf{\textcolor{blue}{\href{#1}{#2}}}}
\newcommand\giturl{\myhref{https://github.com/MilesCranmer/pysr_paper}{github.com/MilesCranmer/pysr\_paper}}

\titlehead{
    \begin{center}
    \parbox{0.3\textwidth}{
        \hfill
        \emph{PySR \& SymbolicRegression.jl}
    }
    \hspace{2in}
    \parbox{0.3\textwidth}{
        \giturl
        \hfill
    }
    \end{center}
}

\title{Interpretable Machine Learning for Science \\ with PySR and SymbolicRegression.jl}
\author[1,2]{Miles Cranmer}
\affil[1]{Princeton University, Princeton, NJ, USA}
\affil[2]{Flatiron Institute, New York, NY, USA}
\date{May 2, 2023}

\twocolumn[\begin{@twocolumnfalse}
\maketitle
\begin{abstract}
\addtocounter{footnote}{2}
\pysr\footnotemark{} is an open-source library for practical \textit{symbolic regression}, a type of machine learning which aims to discover human-interpretable symbolic models.
\pysr was developed to democratize and popularize symbolic regression for the sciences, and is built on a high-performance distributed backend, a flexible search algorithm, and interfaces with several deep learning packages.
\pysr's internal search algorithm is a multi-population evolutionary algorithm, which consists of a unique evolve-simplify-optimize loop, designed for optimization of unknown scalar constants in newly-discovered empirical expressions.
\pysr's backend is an extremely optimized Julia library \pysrjlabs\footnotemark{},
which can be used directly from Julia. It is capable of fusing user-defined operators into SIMD kernels at runtime, performing automatic differentiation, and distributing populations of expressions to thousands of cores across a cluster.
In describing this software, we also introduce a new benchmark, ``\textit{EmpiricalBench},’’ to quantify the applicability of symbolic regression algorithms in science.
This benchmark measures recovery of historical empirical equations from original and synthetic datasets.
\end{abstract}
\end{@twocolumnfalse}
]

\addtocounter{footnote}{-1}
\footnotetext{\textsf{github.com/MilesCranmer/PySR}}
\addtocounter{footnote}{1}
\footnotetext{\textsf{github.com/MilesCranmer/SymbolicRegression.jl}}

\section{Introduction}
\label{sec:intro_pysr}

Johannes Kepler discovered his famous third law of planetary motion, $\text{(period)}^2 \propto \text{(radius)}^3$, from searching for patterns in thirty years of data produced by Tycho Brahe's unaided eye.
Kepler did not discover this by searching through all conceivable relationships with a genetic algorithm on a computer, but by his own geometrical intuition. 
``And it was Kepler's Third Law, not an apple, that led Isaac Newton to discover the law of gravitation''~\cite{hawkingShouldersGiantsGreat2004}.
Likewise, Planck's law was not derived from first principles, but was a symbolic form fit to data~\cite{planckUeberVerbesserungWien1900}.
This symbolic relationship would inspire the development of Quantum Mechanics.

Now, this first step---discovering empirical relationships from data or based on human intuition---is both difficult and time-consuming, even for low-dimensional data where it has been shown to be NP-hard~\cite{virgolinSymbolicRegressionNPhard2022}.
In modern, high-dimensional datasets, it seems an impossible task to discover simple symbolic relationships without the use of automated tools.

This brings us to an optimization problem known as ``symbolic regression,'' or SR.
In this paper, we describe an algorithm for performing SR, and describe software dedicated to equation discovery in the sciences.

\paragraph{Symbolic Regression.}
SR describes a supervised learning task where the model space is spanned by analytic expressions.
This is commonly solved in a multi-objective optimization framework, jointly minimizing prediction error and model complexity.
In this family of algorithms, instead of fitting concrete parameters in some overparameterized general model, one searches the space of simple analytic expressions for accurate and interpretable models.
In the history of science, scientists have often performed SR ``manually,'' using a combination of their intuition and trial-and-error to find simple and accurate empirical expressions.
These empirical expressions then might lead to new theoretical insights, such as the aforementioned discoveries of Kepler and Planck leading to classical and quantum mechanics respectively.
SR algorithms automate this discovery of empirical equations, exploiting the power of modern computing to test many more expressions than our intuition alone could sort through.

SR saw early developments 
as a tool for science beginning in the 1970s and 1980s, with works such as~\cite{gerwinInformationProcessingData1974}, and the creation of the tool \textsc{Bacon} in~\cite{langleyBACONProductionSystem1977} with further development in, e.g.,~\cite{langleyRediscoveringPhysicsBACON1979,langleyBACONDiscoveryConservation1981}.
Via a combination of heuristic strategies for brute force searching over possible expression trees, \textsc{Bacon} and its follow-ups such as \textsc{Fahrenheit}~\cite{langleyDatadrivenApproachesEmpirical1989}, demonstrated the ability to discovery a variety of simple physics laws from idealized data.
The use of genetic algorithms, which allow for a more flexible search space and fewer prior assumptions about these expressions, was popularized with the work of~\cite{kozaGeneticProgrammingMeans1994}.
Finally, with the development of the user-friendly tools \eureqa in~\cite{bongardCoverAutomatedReverse2007,schmidtDistillingFreeFormNatural2009,schmidtSymbolicRegressionImplicit2010}, as well as HeuristicLab in~\cite{wagnerHeuristicLabGenericExtensible2005}, SR started to become practical for use on real-world and noisy scientific data and empirical equations.

\paragraph{Equation Discovery for Science.} 
Today \eureqa is a closed-source proprietary tool, and has fallen out of common use in the sciences.
In addition to other early work such as~\cite{debFastElitistNondominated2000,debFastElitistMultiobjective2002,davidsonSymbolicNumericalRegression2003,cranmerPhysicsGPGeneticProgramming2005},
there have recently been a significant number of exciting new developments in SR as well as SR for science.
These have largely been focused on genetic algorithms on expressions represented as trees, and include core advances to genetic algorithms, such as~\cite{lacavaEpsilonLexicaseSelectionRegression2016,lacavaProbabilisticMultiobjectiveAnalysis2018}, which propose modified strategies for selecting an individual to mutate and~\cite{virgolinScalableGeneticProgramming2017,virgolinImprovingModelbasedGenetic2021} which use an adaptive form of recombination based on emerging statistical patterns.
New strategies for SR altogether have also been developed:
in particular, our previous work~\cite{cranmerLearningSymbolicPhysics2019,cranmerDiscoveringSymbolicModels2020,cranmerDisentangledSparsityNetworks2021} proposes SR as a way of interpreting a neural network.
Neural networks can efficiently find low-dimensional patterns in high-dimensional data, and so this mixed strategy also presents a way of performing SR on high-dimensional or non-tabular data.

Additional techniques include~\cite{jinBayesianSymbolicRegression2020,guimeraBayesianMachineScientist2020} which propose a Markov chain Monte Carlo-like sampling for SR;
\cite{petersenDeepSymbolicRegression2021,liNeuralguidedSymbolicRegression2019} which trains deep generative models to propose mutated expressions;
\cite{dascoliDeepSymbolicRegression2022,kamiennyEndtoendSymbolicRegression2022} which pre-train a large transformer model on billions of synthetic examples, in order to autoregressively generate expressions from data;
\cite{udrescuAIFeynmanParetooptimal2020,liuAIPoincareMachine2020,wetzelDiscoveringSymmetryInvariants2020} which develop techniques for discovering symmetries with SR;
\cite{sahooLearningEquationsExtrapolation2018} which optimizes a sparse set of coefficients in a very large pre-defined expression with stochastic gradient descent; and finally~\cite{atkinsonDatadrivenDiscoveryFreeform2019,rossBenchmarkingMachineLearning2022} which apply genetic algorithms to a flexible space of differential equations.

Apart from trees, a few other data structures are used for representing symbolic expressions, such as imperative representations (``Linear Genetic Programming'')~\citep[e.g.,][]{brameierComparisonLinearGenetic2001,guvenLinearGeneticProgramming2009,maEvolvingSymbolicDensity2022,diasDescribingQuantumInspiredLinear2012}, and sequence representations~\citep[e.g.,][]{liuAIPoincareMachine2020,dascoliDeepSymbolicRegression2022,kamiennyEndtoendSymbolicRegression2022}.
One of the most popular alternatives to genetic algorithms on trees, especially for the study of PDEs, is ``SINDy'' (Sparse Identification of Nonlinear Dynamics) ~\cite{bruntonDiscoveringGoverningEquations2016,rudyDatadrivenDiscoveryPartial2017,championDatadrivenDiscoveryCoordinates2019} which represents expressions as linear combination of a dictionary of fixed nonlinear expressions.
Similar ideas of a linear basis of terms is used in the initialization of the ``Operon'' genetic algorithm~\cite{burlacuOperonEfficientGenetic2020}, as well as in ``FFX''~\cite{mcconaghyFFXFastScalable2011}.
This strategy can also be combined with deep learning~\citep[e.g., ][]{luschDeepLearningUniversal2018,championDatadrivenDiscoveryCoordinates2019,bothDeepMoDDeepLearning2019,chenDeepLearningPhysical2020,rackauckasUniversalDifferentialEquations2020} or other techniques to find expressions on high-dimensional data~\citep[e.g., ][]{rossBenchmarkingMachineLearning2022}.
Additional techniques include mixed integer nonlinear programming~\citep{cozadGlobalMINLPApproach2018} and greedy tree searches~\citep{defrancaGreedySearchTree2018}, and combining theory as a type of constraint over the space of models~\cite{cornelioAIDescartesCombining2021}.

However, here we argue that many existing algorithms lack certain features which make them practical in a broad setting for discovering interpretable equations in science.
The requirements for ``scientific SR'' are significantly more difficult than for SR applied to synthetic datasets, which many algorithms are implicitly trained on via synthetic benchmarks.
We list these below.

\begin{easylist}
\ListProperties(Margin1=1cm,Hang1=true,Indent1=.5cm,Numbers1=a,Start1=1,Align=0.5cm)
#
an empirical equation in science quite often contains unknown real constants\footnote{One could argue that any constant in a scientific model must depend on a finite number of known constants of physics (e.g., the speed of light) and mathematics (e.g., $\pi$). However, this introduces significantly more complexity than is being reduced, as the relationship with fundamental constants to derived constants might be arbitrarily complex.}
The set of possible expressions considered by some SR algorithms such as~\cite{udrescuAIFeynmanParetooptimal2020}: operators, integers, and variables---is discrete and finite up to a given max size.
This set is even searchable with brute force, as is used in~\cite{udrescuAIFeynmanParetooptimal2020} in combination with a symmetry and dimensional analysis module.
However, with arbitrary real constants, the set of equations one must search is uncountably infinite.
The difficulty of searching this set is further compounded by several factors:
#
The discovered equation must provide insight into the nature of the problem.
The equation which often holds the most insight, and is therefore adopted by scientists, is not always the most accurate (many other ML algorithms could be used instead for accuracy), but is instead an equation which balances accuracy with simplicity.
#
An empirical equation in science is discovered from data containing noise, which often has heteroscedastic structure.
#
Many empirical equations are not differentiable, instead being composed of different expressions active in only parts of parameter space (e.g., $\exp(x)$ for $x<0$ and $x^2+1$ for $x\geq 0$).
#
Discovered expressions in science must satisfy known constraints, e.g., mass conservation.
#
Empirical relations are frequently comprised of operators which are unique to one particular field of science, so the equation search strategy must allow for custom operators.
#
Observational data is often high-dimensional, and the SR algorithm must either manage this internally, or be layered with a feature selection algorithm.
#
The search tool must be capable of finding relationships in non-tabular datasets, such as sequences, grids, and graphs (for example, by using aggregation operators such as $\sum$). #
Finally, and most importantly, any SR package useful to scientists must be open-source, cross-platform, easy-to-use, and interface with common tools.
For more discussion, the references~\cite{price-whelanAstropyProjectBuilding2018,astropycollaborationAstropyProjectSustaining2022} consider this from the point of view of the astrophysics community.
\end{easylist}

\paragraph{Contribution.}
In this paper, we introduce \pysr, a Python and Julia package for Scientific Symbolic Regression. The Python package is available under \texttt{pysr} on PyPI and Conda\footnote{\myhref{https://github.com/MilesCranmer/PySR}{github.com/MilesCranmer/PySR}} and the Julia library under \texttt{SymbolicRegression}\footnote{\myhref{https://github.com/MilesCranmer/SymbolicRegression.jl}{github.com/MilesCranmer/SymbolicRegression.jl}} on the Julia package registry.
Every part of \pysr is written with scientific discovery in mind, as the entire point of creating this package was to enable the authors of this paper to use this tool in their own field for equation discovery.
All of the requirements of discovering an empirical equation for science, as listed above, are satisfied by \pysr.
We discuss this in detail in the following sections.

\section{Methods}
\label{sec:methods}

First, we give an in-depth discussion of the internal algorithm in \pysr in \cref{sec:algorithm}, and then describe the software implementation in \cref{sec:software}.
We then discuss a few additional implementation features in \cref{sec:interfaces}, \cref{sec:custom_operators}, and \cref{sec:custom_losses}.

\subsection{Algorithm}
\label{sec:algorithm}

\pysr is a multi-population evolutionary algorithm with multiple evolutions performed asynchronously.
The main loop of \pysr, operating on each population independently, is a classic evolutionary algorithm based on tournament selection for individual selection (first introduced in the thesis of~\cite{brindleGeneticAlgorithmsFunction1980} and generalized in~\cite{goldbergComparativeAnalysisSelection1991}),
and several mutations and crossovers for generating new individuals.
Evolutionary algorithms for SR was first popularized in~\cite{kozaGeneticProgrammingMeans1994}, and has dominated the development of flexible SR algorithms since.
\pysr makes several modifications to this classic approach, some of which are motivated by results in recent work.

\begin{figure}[h!]
    \centering
    \includegraphics[page=3,width=0.9\linewidth]{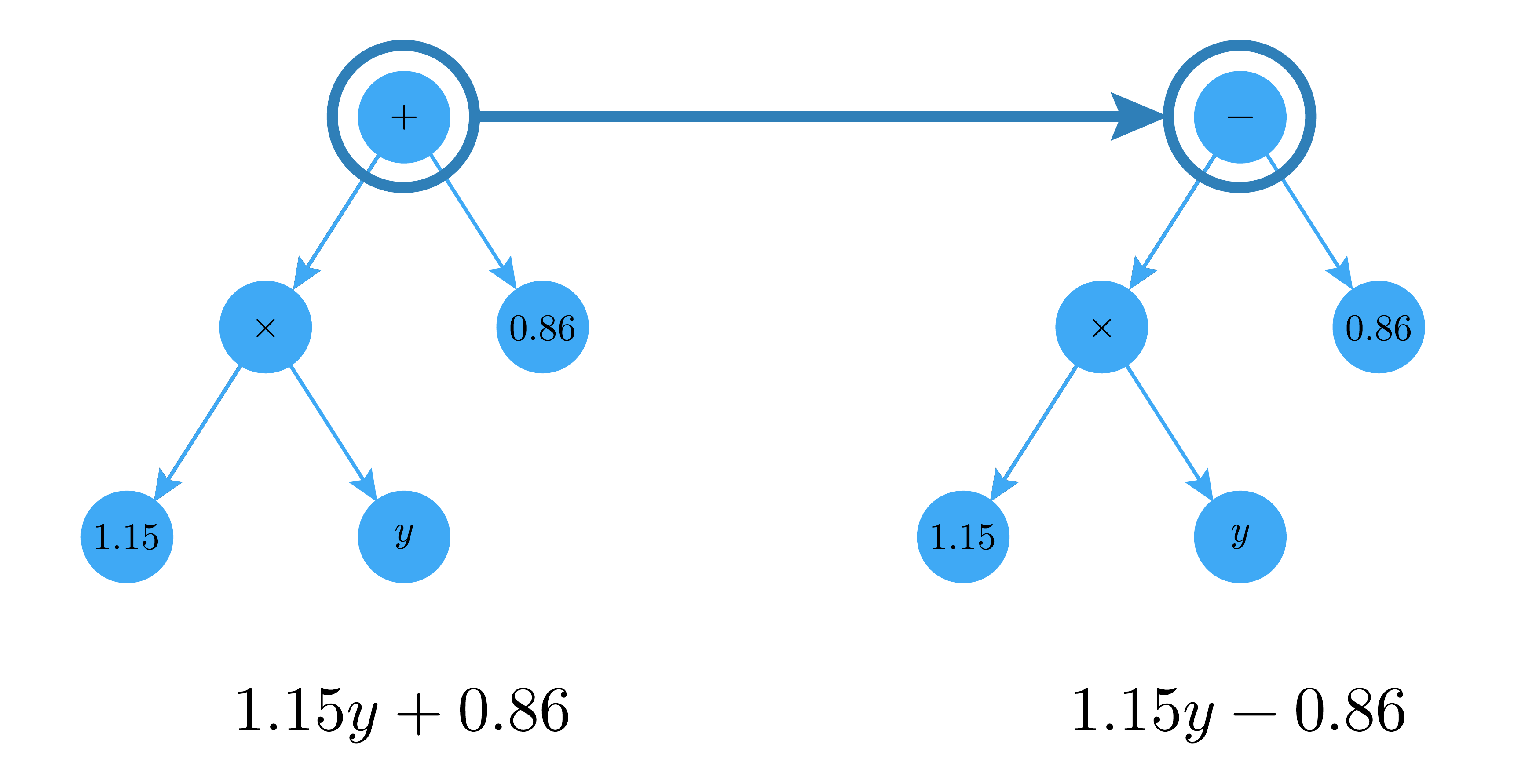}
    \caption{A mutation operation applied to an expression tree.}
    \label{fig:mutations}
\end{figure}

\begin{figure}[h!]
    \centering
    \includegraphics[page=5,width=0.9\linewidth]{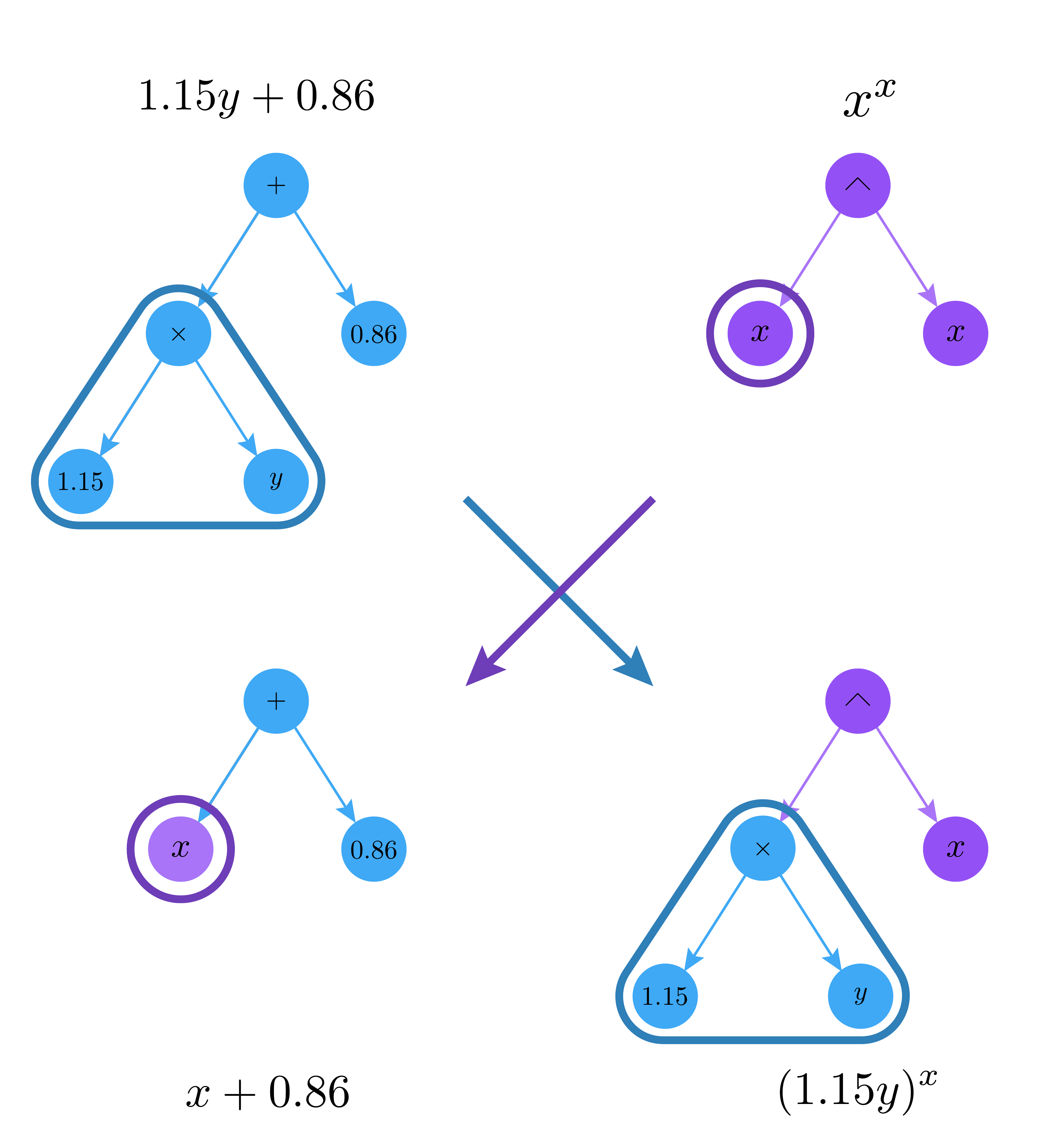}
    \caption{A crossover operation between two expression trees.}
    \label{fig:crossover}
\end{figure}

\begin{figure*}[h!]
    \centering
    \includegraphics[page=2,width=0.7\linewidth]{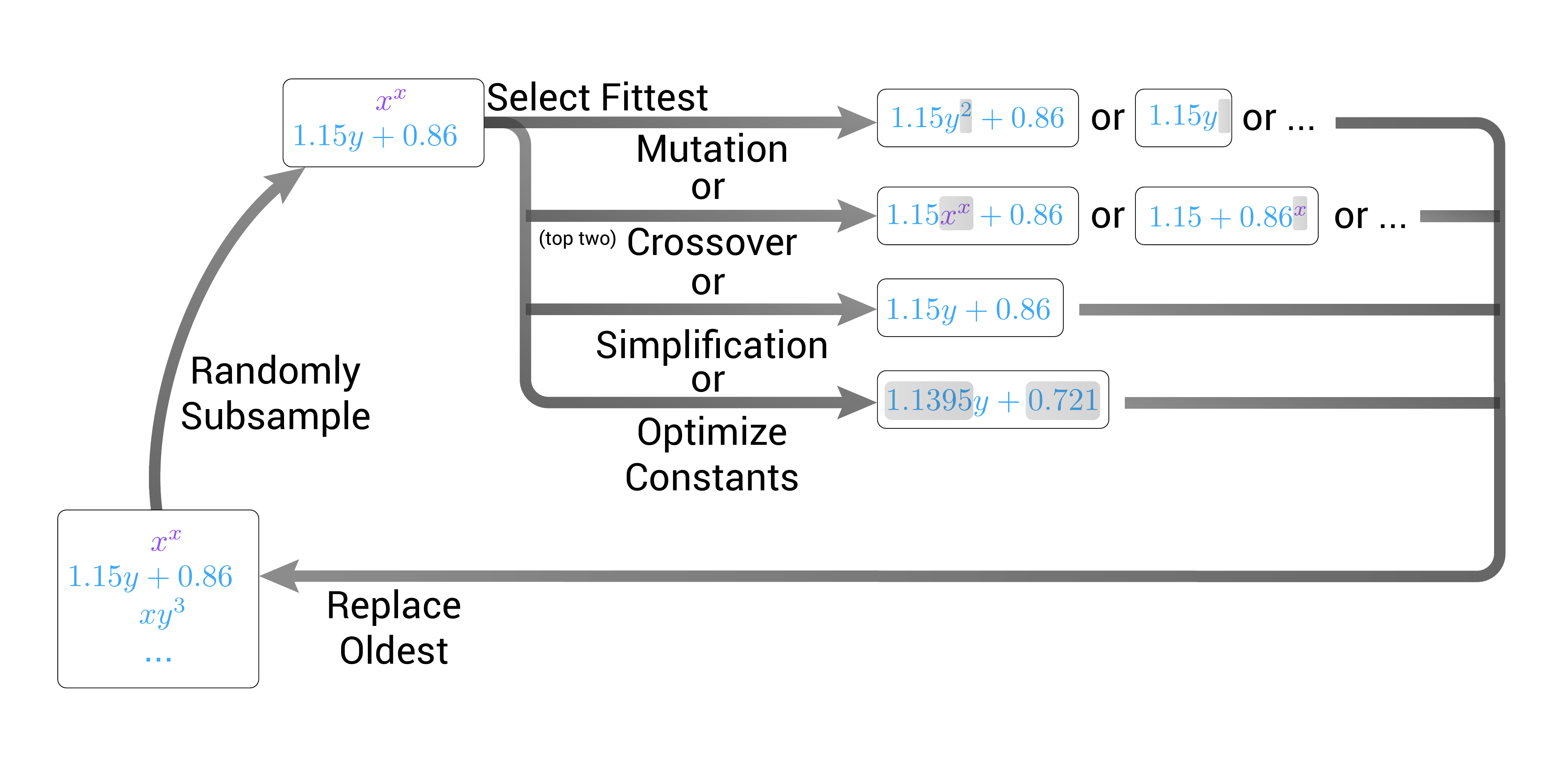}
    \caption{The inner loop of \pysr. A population of expressions is randomly subsampled. Among this subsample, a tournament is performed, and the winner is selected for breeding: either by mutation, crossover, simplification, or explicit optimization. Examples of mutation and crossover operations are visualized in \cref{fig:mutations,fig:crossover}.}
    \label{fig:evolution_loop}
\end{figure*}

\begin{figure*}[h!]
    \centering
    \includegraphics[page=1,width=0.7\linewidth]{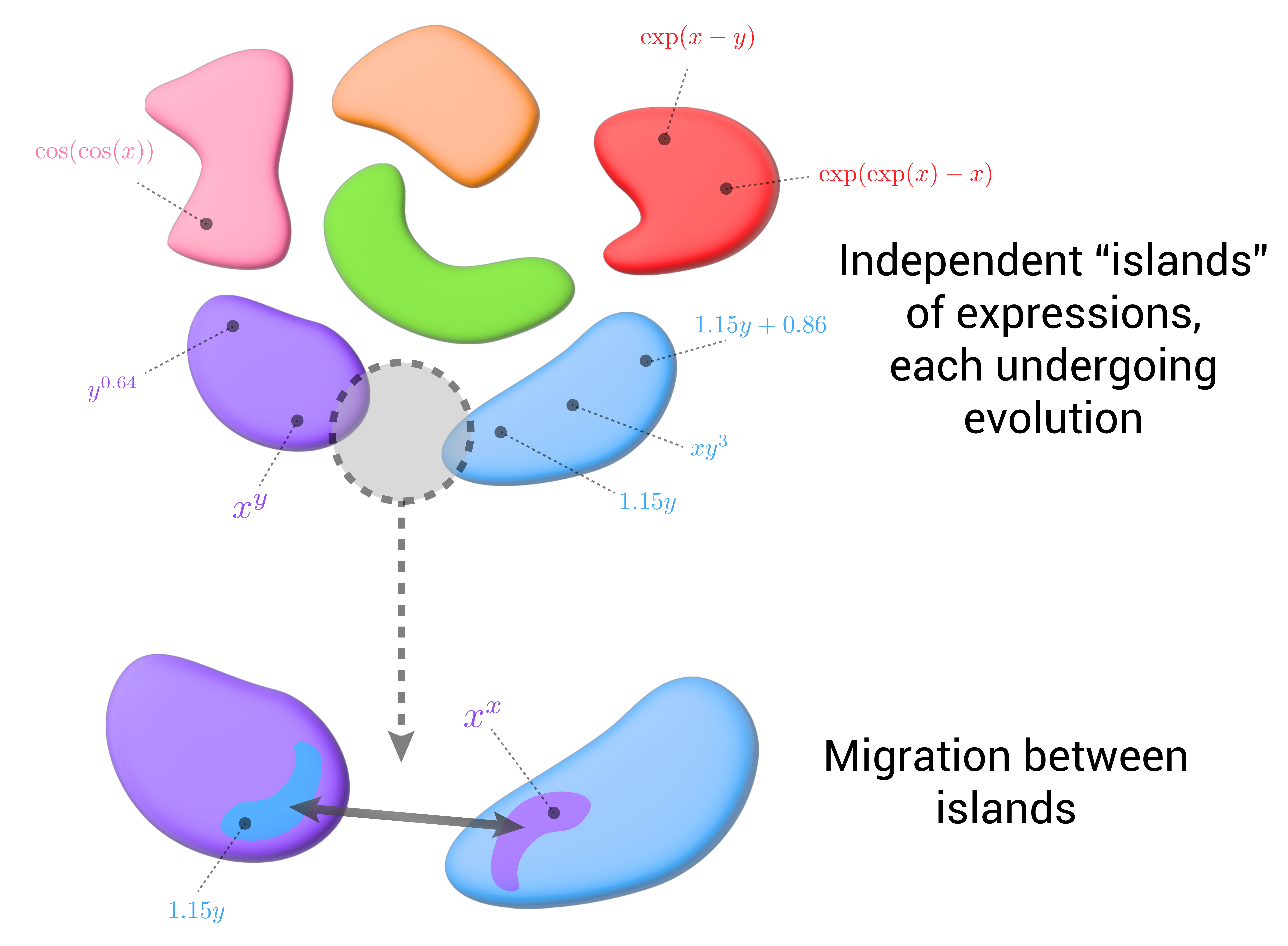}
    \caption{The outer loop of \pysr. Several populations evolve independently according to the algorithm described in \cref{fig:evolution_loop}. At the end of a specified number of rounds of evolution, migration between islands is performed.}
    \label{fig:migration}
\end{figure*}

A simple evolutionary algorithm proceeds as follows:
\begin{easylist}
\ListProperties(Margin1=1cm,Hang1=true,Indent1=.5cm,Numbers1=a,Start1=1,Align=0.5cm)
# Assume one has a population of individuals, a fitness function, and a set of mutation operators.
\label{step:assumptions}
# Randomly select an $n_s$-sized subset of individuals from the population (e.g., classical tournament selection uses $n_s=2$, but larger tournaments are also allowed).
\label{step:sampling}
# Run the tournament by evaluating the fitness of every individual in that subset.
\label{step:prep_tournament}
# Select the fittest individual as the winner with probability $p$. Otherwise, remove this individual from the subset and repeat this step again.
If there is one remaining, select it.
Thus, the probability will roughly be $p, p(1-p), p(1-p)^2, \ldots$ for the first, second, third fittest, and so on. 
\label{step:select}
# Create a copy of this selected individual, and apply a randomly-selected mutation from a set of possible mutations.
\label{step:mutate}
# Replace a member of the population with the mutated individual. Here, one would replace the weakest of the population or subset.
\label{step:replace}
\end{easylist}

This algorithm is computationally efficient and allows for massive parallelization by splitting step~\ref{step:sampling} into groups of individuals which can evolve independently, with asynchronous ``migration'' between the groups.
For the first of the modifications, we replace the eldest member of a population in step~\ref{step:mutate}, rather than the individual with the lowest fitness, which is known as ``age-regularized'' evolution~\cite{realRegularizedEvolutionImage2019}.
We record the age from the Unix time at which the individual was created during step~\ref{step:mutate}.
Its inclusion in \pysr was motivated by the impressive Neural Architecture Search results of~\cite{realRegularizedEvolutionImage2019,realAutoMLZeroEvolvingMachine2020a} --- though it is worth noting similar regularizations used earlier in SR, e.g.,~\cite{hornbyALPSAgelayeredPopulation2006} and~\cite{schmidtAgefitnessParetoOptimization2010} --- which found this simple modification could prevent early convergence without hurting performance.
This prevents the population from specializing too quickly and getting stuck in a local minimum of the search space.
The implementation in~\cite{realAutoMLZeroEvolvingMachine2020a} uses $p=1$ and explores $4\leq n_s \leq 64$, though here we allow $p$ to be less than $1$, among several other differences.

\paragraph{Modifications.}
Our algorithm in \pysr, first released publicly in 2020~\cite{cranmerPySRFastParallelized2020a}, makes three changes to this evolutionary algorithm.
We apply simulated annealing~\cite{kirkpatrickOptimizationSimulatedAnnealing1983} to step~\ref{step:select}: given a temperature $T\in[0, 1]$, a mutation is rejected with some probability related to the change in fitness after mutation.
The probability for rejection is $p=\exp(\frac{L_F - L_E}{\alpha T})$, for $L_F$ and $L_E$ the fitness of the mutated and original individual respectively, and $\alpha$ a hyperparameter.
This allows the evolution to alternate between \textit{high temperature} and \textit{low temperature} phases, with the high temperature phases increasing diversity of individuals and the low temperature phases narrowing in on the fittest individuals.
$\alpha$ controls the scale of temperature, with $\alpha\rightarrow\infty$ being equivalent to regular tournament selection, and $\alpha\rightarrow 0$ rejecting all mutations which lower fitness.
Since the value of $p$ also controls the diversity of the population, we can also couple the value of $p$ to temperature, setting a larger value of $p$ for a higher temperature, and a smaller value of $p$ for a lower temperature.
Simulated annealing is a simple yet powerful strategy for optimizing over discrete spaces, and we find in \cref{sec:evaluation} that simulated annealing can significantly speed up the search process.

The second modification we make to tournament selection is specific to the individuals we wish to consider: mathematical expressions.
We embed the genetic algorithm inside an \textit{evolve-simplify-optimize loop}.
``Evolve'' is a repeated application of tournament selection-based evolution for a set number of mutations.
In other words, this is an entire evolution of a population.
``Simplify'' here refers to equation simplification---equations are simplified to an equivalent form during the loop using a set of algebraic equivalencies (since this happens infrequently enough, it tends to not harm discovery by constraining the search space).
The ``Optimize'' part of this algorithm is a few iterations of a classical optimization algorithm (BFGS~\cite{broydenConvergenceClassDoublerank1970} is used by default, though any optimizer in the \texttt{Optim.jl} package can be used~\cite{mogensenOptimMathematicalOptimization2018}) to optimize constants in every equation explicitly.
This part of the algorithm significantly improves the discovery of equations containing real constants, which is essential for practical application of SR to science.
It has been shown since at least~\cite{topchyFasterGeneticProgramming2001} that performing local gradient searches on numerical constants in expressions can vastly improve the performance of SR---we closely integrate an optimized and stochastic version of this.

The reason why several mutations (as well as crossovers) are performed before proceeding to the \textit{simplify-optimize} stage is because some equations are only accessible via a redundant intermediate state.
For example, $x\ast x - x\ast x$ would normally simplify to $0$, but if the simplification is avoided, the next mutation might change one of the $x$'s to $y$, resulting in $x\ast x - x \ast y$.
Simplifying only occasionally is a way of allowing redundant but necessary intermediate expressions while at the same time reducing the size of the search space.

Finally, the third modification we make is the use of a novel \textit{adaptive parsimony} metric.
But first, we give a definition of complexity.

\def\boxmargins{12pt}
\newcommand{\greybox}[1]{
\begin{center}
    \tcbox[
        nobeforeafter,
        on line,
        boxrule=0.25pt,
        colback=gray!20,
        colframe=gray!50,
        arc=3pt,
        left=\boxmargins,
        right=\boxmargins,
        top=\boxmargins,
        bottom=\boxmargins,
        boxsep=0pt,
        width=\linewidth-\boxmargins-25pt
    ]{
        \parbox{\linewidth-\boxmargins-30pt}{\small #1}
    }
\end{center}
}
\greybox{
    \textbf{Complexity}:
    By default, complexity in \pysr is equal to the \textit{number of nodes in an expression tree}, regardless of each node's content.

    However, complexity in \pysr is completely user-configurable. We argue that the best definition for a ``simple'' expression is the one that is most interpretable by the user. An expression containing a dilogarithm may look commonplace to the particle physicist---and in fact be a great prior over the space of models for certain particle physics data---but extremely unusual to the fluid dynamicist, and perhaps not coincidentally a bad choice of prior for fluids data.
}

\noindent The traditional mechanism for penalizing complexity would be to use a constant \textit{parsimony}, which would involve adding a loss term equal to the complexity of an expression times a constant factor:
\begin{equation}
    \ell(E) = \ell_\text{pred}(E) + (\text{parsimony}) \cdot C(E),
\end{equation}
for $C(E)$ the complexity of an expression $E$ and $\ell_\text{pred}$ the predictive loss.
Instead, we adaptively tune a per-complexity penalty such that the number of expressions at each complexity is approximately the same.
This is expressed roughly as:
\begin{equation}
    \ell(E) = \ell_\text{pred}(E) \cdot \exp( \text{frecency}[C(E)]),
\end{equation}
where the ``frecency'' of $C(E)$ is some combined measure of the frequency and recency of expressions occurring at complexity $C(E)$ in the population.
In \pysr, we simply count the number of expressions with a moving window in time, and divide by a tunable constant.
Thus, this is a simple way to punish complexities adaptively by how frequent they are in the population.
This encourages the evolutionary search to explore the problem from simple, less-accurate expressions, as well as from complex, more accurate expressions.
This qualitatively seems to help alleviate situations where the search specializes too early in the wrong functional form, and can only seem to make small iterations to it, and is discouraged from starting from scratch.

Pseudocode for the outer loop of \pysr is given in \cref{alg:pysr}, which includes the migration steps shown in \cref{fig:migration}.
\Cref{alg:evolve} gives pseudocode for the main \textit{evolve-simplify-optimize loop}, formalizing parts of the cartoon in \cref{fig:evolution_loop}.
\Cref{alg:mutate} outlines the set of mutations, supplemented with simulated annealing.
Finally, \cref{alg:tournament} describes the tournament selection strategy.

\newcommand\algorithmsize{\small}

\newcommand\alggray{40}
\newcommand\algcomment[1]{{\itshape\color{commentsColor}// #1}}

\newcommand\setinsert[3]{(#1 \setminus #2) \cup #3}
\newcommand\loopdef{\gets }
\newcommand\npop{n_\text{p}}
\newcommand\nper{L}
\newcommand\ncyclesper{n_\text{c}}
\newcommand\options{\text{options}}
\newcommand\data{\mathcal{X}}
\newcommand\atext[1]{\textsc{#1}}
\newcommand\rand{\atext{Rand}()}
\newcommand\ton{$1$ \KwTo $\npop$}
\newcommand\niter{n_\text{iter}}

\SetStartEndCondition{ }{}{}\SetKwProg{Fn}{function}{}{end}
\SetKwFunction{Range}{range}\SetKwFunction{Loss}{loss}
\SetKwFunction{Copy}{copy}
\SetKwFunction{Evolve}{evolve}\SetKwFunction{Simplify}{simplify}\SetKwFunction{Rand}{rand}\SetKwFunction{RandChoice}{rand\_choice}\SetKwFunction{Randn}{randn}\SetKwFunction{pysrfnc}{pysr}\SetKwFunction{Tournament}{tournament}\SetKwFunction{GetFittest}{get\_fittest}\SetKwFunction{Mutate}{mutate}\SetKwFunction{TakeBest}{take\_best}\SetKwFunction{SatisfiesConstraints}{satisfies\_constraints}\SetKw{KwTo}{in}
\SetKwFor{For}{for}{}{end}\SetKwIF{If}{ElseIf}{Else}{if}{}{else if}{else}{end}\SetKwFor{While}{while}{}{end}\SetKwRepeat{Repeat}{}{}\SetKwSwitch{Switch}{Case}{Other}{switch}{}{case}{otherwise}{}{end}\SetKwInOut{Param}{param}
\SetKwInOut{Data}{input}
\SetKwInOut{Result}{output}
\SetKwFunction{Null}{}

\newcommand\algsize\scriptsize
\setlength{\algomargin}{1.5em}

\begin{algorithm*}[h!]
\algsize
\DontPrintSemicolon
\Data{$\data$, the dataset to find expressions for}
\Result{the best expressions at each complexity}
\Param{$n_p$, the number of populations (=40)}
\Param{$\nper$, the number of expressions in each population (=1000)}
\Param{$\alpha_H$, replacement fraction using expressions from $H$ (=0.05)}
\Param{$\alpha_M$, replacement fraction using expressions from $\cup_i M_i$ (=0.05)}
\BlankLine
\Fn{\pysrfnc{$\data$}}{
\For{i \KwTo \Range{$\npop$} \algcomment{\href{https://github.com/MilesCranmer/SymbolicRegression.jl/blob/0bb8b2bb58a4338729c84e3f52575ac475c86a09/src/SymbolicRegression.jl\#L534-L546}{[code]}}}{
create set $P_i$ containing $\nper$ random expressions of complexity 3\;
\algcomment{\^ e.g., $(3.2 + x_1)$ has size 3}\;
create empty set $M_i$ \algcomment{will store best expressions seen in $P_i$}\;
}
create empty set $H$ \algcomment{will store best expressions overall}\;
\For{$n$ \KwTo \Range{$\niter$} \algcomment{\href{https://github.com/MilesCranmer/SymbolicRegression.jl/blob/0bb8b2bb58a4338729c84e3f52575ac475c86a09/src/SymbolicRegression.jl\#L674-L1055}{[code]}}}{
    \algcomment{the following loop is parallelized over workers:}\;
    \For{$i$ \KwTo \Range{$\npop$}}{
        \BlankLine
        \algcomment{evolve-simplify-optimize:}\;
        $P_i \gets $ \Evolve{$P_i, \data$}\;
        \For{$E$ \KwTo $P_i$}{
            simplify $E$\;
            optimize constants in $E$\;
            store updated $E$ in $P_i$
        }
        $M_i\gets$ most accurate expression in $P_i$ at each complexity\;
        \algcomment{(In actuality, $M_i$ is updated throughout \texttt{evolve})}\;
        $H\gets$ most accurate expression in $M_i \cup H$ at each complexity\;
        \BlankLine
        \algcomment{migration:}\;
        \For{$E$ \KwTo $P_i$}{
            \If{\Rand{} $<\alpha_H$}{
                replace $E$ in $P_i$ with a random expression from $H$
            }
            \If{\Rand{} $<\alpha_M$}{
                replace $E$ in $P_i$ with a random expression from $\cup_{j\neq i} M_j$
            }
        }
    }
}
\KwRet{$H$}
}
\caption{PySR\label{alg:pysr}.}
\end{algorithm*}

\begin{algorithm}[h!]
\algsize
\DontPrintSemicolon
\Data{$P$, a set of expressions}
\Data{$\data$ as in \cref{alg:pysr}}
\Result{$P$, the evolved set of expressions}
\Param{$n_c$, the number of mutations per \Evolve{} call (=300000)}
\Param{$p_\text{cross}$, the probability of crossover (=0.01)}
\BlankLine
\Fn{\Evolve{$P, \data$}}{
\For{$k$ \KwTo \Range{$n_c$} \algcomment{\href{https://github.com/MilesCranmer/SymbolicRegression.jl/blob/0bb8b2bb58a4338729c84e3f52575ac475c86a09/src/SingleIteration.jl\#L36-L58}{[code]}}}{
    \eIf{$\Rand{} > p_\text{cross}$ \algcomment{\href{https://github.com/MilesCranmer/SymbolicRegression.jl/blob/0bb8b2bb58a4338729c84e3f52575ac475c86a09/src/RegularizedEvolution.jl\#L81-L154}{[code]}}}{
        \algcomment{mutation}\;
        $E \gets $ \Tournament{$P, \data$}\;
        $T\gets 1 - \frac{k}{n_c}$ \algcomment{annealing temperature}\;
        $E^\ast \gets $ \Mutate{$E,$ T}\;
        replace oldest expression in $P$ with $E^\ast$
    }{
        \SetKwRepeat{Do}{do}{until}
\Do{\textnormal{\SatisfiesConstraints{$E^\ast_1$}} and \\ \ \ \textnormal{\SatisfiesConstraints{$E^\ast_2$}}}{
            \algcomment{crossover}\;
            $E_1 \gets $ \Tournament{$P, \data$}\;
            $E_2 \gets $ \Tournament{$P, \data$}\;
            \SetKwFunction{Crossover}{crossover}
            $E_1^\ast, E_2^\ast \gets $ \Crossover{$E_1, E_2$}\;
            replace oldest two expressions in $P$ with $E_1^\ast$ and $E_2^\ast$
        }
    }
}
\KwRet{$P$}
}
\caption{Evolution.\label{alg:evolve}}
\end{algorithm}

\begin{algorithm}[h!]
\algsize
\DontPrintSemicolon
\Data{$E$, an expression}
\Data{$T$, the annealing temperature $\in [0, 1]$}
\Result{mutated version of $E$}
\Param{$m_i$, for $i=1,\ldots,8$, the probability weight of mutation type $i$}
\Param{$f$, the constant perturbation scale (=1)}
\Param{$\epsilon$, the minimum perturbation (=0.1)}
\Param{$\alpha$, the temperature scale (=0.1)}
\BlankLine
\Fn{\Mutate{$E, T$} \algcomment{\href{https://github.com/MilesCranmer/SymbolicRegression.jl/blob/0bb8b2bb58a4338729c84e3f52575ac475c86a09/src/Mutate.jl\#L22-L261}{[code]}}}{
adjust weights $w_1, \ldots, w_8$ based on constraints\;
$i \gets$ random choice of $ 1, \ldots, 8$ weighted by $w_1, \ldots, w_8$\;
\SetKwRepeat{Do}{do}{until}
\Do{\textnormal{\SatisfiesConstraints{$E^\ast$}}}{
    $E^\ast \gets $ \Copy{$E$}\;
    \Switch{$i$}{
        \Case(\algcomment{mutate constant}){$1$}{
            $a \gets  (1 + f \times T + \epsilon)^{2\times \Rand{} - 1}$\;
            \If{\Rand{} $< 0.5$}{
                $a \gets -a$
            }
            multiply random constant in $E^\ast$ by $a$
        }
        \Case(\algcomment{mutate operator}){$2$}{
            randomly replace an operator in $E^\ast$ with\\
            \ \ an operator of the same degree
        }
        \Case(\algcomment{append/prepend node}){$3$}{
            add random node to root or leaf of $E^\ast$
        }
        \Case(\algcomment{insert node}){$4$}{
            insert a random node inside $E^\ast$
        }
        \Case(\algcomment{delete subtree}){$5$}{
            replace a node and its children from $E^\ast$\\
            \ \ with a constant or variable
        }
        \Case(\algcomment{simplify tree}){$6$}{
            simplify $E^\ast$
        }
        \Case(\algcomment{new tree entirely}){$7$}{
            $E^\ast \gets $ random expression
        }
        \Case(\algcomment{no mutation}){$8$}{
            do nothing
        }
    }
}
\BlankLine
$C, C^\ast \gets$ complexity of expressions $E, E^\ast$, respectively\;
$L, L^\ast \gets$ accuracy of expressions $E, E^\ast$, respectively\;
$q_\text{anneal} \gets \exp(-\frac{L^\ast - L}{\alpha \times T})$\;
$C \gets$ complexity of expressions $E$\;
$q_\text{parsimony} \gets \frac{ \text{frecency}[C] }{ \text{frecency}[C]}$\;
\eIf{\Rand{} $< q_\text{anneal} \cdot q_\text{parsimony}$}{
    \KwRet{$E^\ast$}
}{
    \KwRet{$E$}
}
}
\caption{Mutations.\label{alg:mutate} }
\end{algorithm}

\SetKwFunction{Length}{length}
\SetKw{Break}{break}

\begin{algorithm}[h!]
\algsize
\DontPrintSemicolon
\Data{$P$, a population of expressions}
\Data{$\data$ as in \cref{alg:pysr}}
\Result{a single expression (the winner of the tournament)}
\Param{$n_s$, the tournament size (=12)}
\Param{$p_\text{tournament}$, the probability of selecting the fittest individual (=0.9)}
\BlankLine
\Fn{\Tournament{$P, \data$} \algcomment{\href{https://github.com/MilesCranmer/SymbolicRegression.jl/blob/0bb8b2bb58a4338729c84e3f52575ac475c86a09/src/Population.jl\#L100-L121}{[code]}}}{
    $Q \gets $ a random subset of size $n_s$ of $P$\;
    \While{$\Length{Q} > 1$}{
        {$E \gets \GetFittest{Q}$}\;
        \If{$\Rand{} < p_\text{tournament}$}{
            \Break
        }
        remove $E$ from $Q$
    }
    \KwRet{$E$}
}
\BlankLine
\Fn{\GetFittest{$P$} \algcomment{\href{https://github.com/MilesCranmer/SymbolicRegression.jl/blob/0bb8b2bb58a4338729c84e3f52575ac475c86a09/src/Population.jl\#L86-L95}{[code]}}}{
    $\ell_\text{best} \gets \infty$ \;
    \For{$E$ \KwTo $P$}{
        $C \gets$ complexity of $E$\;
        $\ell \gets \text{accuracy of }E$ \;
        \BlankLine
        \algcomment{include adaptive parsimony:}\;
        ${\ell \gets \ell}\times\exp(\text{frecency}[C])$\;
        \BlankLine
        \If{$\ell < \ell_\text{best}$}{
            $\ell_\text{best} \gets \ell$ \;
            $E^\ast \gets E$
        }
    }
    \KwRet{$E^\ast$}
}
\caption{Tournament selection.\label{alg:tournament} }
\end{algorithm}

\paragraph{Additional Features.}
\pysr includes a variety of additional features for various types of data and scenarios:

\begin{easylist}[itemize]
\ListProperties(Margin1=0.5cm,Margin2=0.8cm)
# \textit{Noisy data.} \pysr supports multiple strategies for working with noisy data:
## First, \pysr includes an optional denoising preprocessing step that optimizes a Gaussian process on the input dataset with a kernel of the form $k(x,x')=\sigma^2\exp(-\abs{x-x'}^2 / 2 l^2) + \alpha\;\delta(x-x') + C$, a superposition of a Gaussian kernel (the standard kernel for interpolating datasets), a white noise kernel (to account for intrinsic noise), and a constant kernel (which modifies the mean of the Gaussian process).
After this kernel is optimized, the Gaussian process is used to predict denoised targets for each input point, which are then passed to the main \pysr algorithm.

## Second, \pysr allows one to specify a set of weights for each input data point. This could be used, for instance, if a user knows the uncertainty in a measurement beforehand to be $\sigma>0$, for $\sigma$ the standard deviation of the measurement, and thus weights each data point in the loss by the signal-to-noise $1/\sigma^2$.

## Thirdly, the user can define a custom likelihood to optimize, which can also take into account weights. This is explained in the following point.
# \textit{Custom losses.}
\pysr supports arbitrary user-defined loss functions, passed as either a string or a function object.
These take a scalar prediction and target, and optionally a scalar weight, and should return a real number greater than 0.
These per-point losses are then summed over the dataset.
With this, the user is capable of defining arbitrary regression objectives, custom likelihoods (defined as log-likelihood of a single data point), and classification-based losses.
This also allows one to define implicit objectives.
# \textit{Custom operators.}
One of the most powerful features of \pysr is the ability to define custom operators.
Different domains of science have functions unique to their field which appear in many formulae and hold a specific meaning.
It is therefore very important that these operators be available in an SR library, and so \pysr allows custom user-defined unary or binary functions.
So long as a function can be defined as either
$f:\mathbb{R}\rightarrow \mathbb{R}$ or $\mathbb{R}^2\rightarrow\mathbb{R}$, it can be used as a user-defined operator.
\pysr \ul{does not treat built-in operators any differently than user-defined ones.}
Apart from simplification strategies, \pysr does not know the difference between the $+$ operator and a Bessel function.
# \textit{Feature selection.}
Similar to the Gaussian process preprocessing step for denoising, \pysr also uses a simple dimensionality reduction preprocessing step.
Given a user-defined number of features to select, \pysr uses a gradient-boosting tree algorithm to first fit the dataset, then select the most important features.
These features are fed into the main search loop.
# \textit{Constraints.}
\pysr allows various hard constraints to be specified for discovered expressions.
These are enforced at every mutation: if a constraint is violated, the mutation is rejected. A few are available with the API:
## The maximum size of an expression (the number of instances of operators, variables, and constants), which gives an overall bound on the search space.
## The maximum depth of an expression, which can be used to control how deeply nested the resultant expression is.
## The maximum size of a subexpression in a specific operator.
This specifies the maximum size of an expression within an operator's arguments.
For example, specifying that the $\wedge$ operator has maximum argument size of $(-1, 3)$ means that its base can have any size expression (indicated by -1), while the expression in its exponent can only have max size of up to 3.
This operator-specific constraint can drastically reduce the complexity of discovered expressions.
## The maximum number of nests of a particular operator combination. For example, the nested constraint \texttt{\{sin: \{sin: 0, cos: 1\}\}} would indicate that $\sin(\cos(x))$ is allowed, but $\sin(\cos(\cos(x)))$ and $\sin(\sin(x))$ are not.
# \textit{Additional data structures}. See deep learning interface \cref{sec:interfaces}.
# \textit{Integrals and dynamical systems}. See deep learning interface \cref{sec:interfaces}.
\end{easylist}

\subsection{Software implementation}
\label{sec:software}

The search algorithm itself underlying \pysr, as described in pseudocode in \cref{alg:pysr,alg:evolve,alg:mutate,alg:tournament}, is written in pure-Julia under the library name \pysrjl\footnote{\myhref{github.com/MilesCranmer/SymbolicRegression.jl}{https://github.com/MilesCranmer/SymbolicRegression.jl}}.
Julia boasts a combination of a high-level interface\footnote{\myhref{https://docs.julialang.org/en/v1/}{docs.julialang.org/en/v1/}} with very high performance comparable to languages such as C++ and Rust\footnote{\myhref{https://julialang.org/benchmarks/}{julialang.org/benchmarks/}}.
However, the key advantage of using Julia for this work is the fact that it is a just-in-time compiled (JIT) language.
This allows \pysr to make use of optimizations not possible with a statically compiled library: for instance, compiling user-defined operators and losses into the core functions.
The most significant advantage of using JIT compilation for \pysr, in terms of performance, is that operators can be fused together into single compiled operators.
For example, if a user declares \texttt{+} and \texttt{-} as valid operators, then \pysr will compile a SIMD-enabled kernel for \texttt{+}, \texttt{-}, as well as their combination: \texttt{(a - b) + c}, and so on. This happens automatically for every single combination of operators up to a depth of two operators, even for user-defined operators.
Simply by fusing operators in this way, the expression evaluation code, which remains the bottleneck in \pysr, experiences a significant speedup.
Because Julia is JIT compiled, these operators need not be pre-defined in the library: they will be just as performant as if they were.

\pysr exposes a simple Python frontend API to this pure-Julia
backend search algorithm.
This frontend takes the popular style
of the \texttt{scikit-learn} machine learning library~\cite{pedregosaScikitlearnMachineLearning2011b}:
\begin{minipage}{0.95\linewidth}
\begin{minted}{python}
from pysr import PySRRegressor

# Declare search options:
model = PySRRegressor(
  model_selection="best",
  unary_operators=["cos", "sin"],
  binary_operators=["+", "-", "/", "*"],
)

# Load the data
X, y = load_data()
# X shape: (n_rows, n_features)
# y shape: (n_rows) or (n_rows, n_targets)

# Run the search:
model.fit(X, y)

# View the discovered expressions:
print(model)

# Evaluate, using the 5th expression along
# the Pareto front:
y_predicted = model.predict(X, 5)
# (Without specify `5`, it will select an expression
# which balances complexity and error)
\end{minted}
\end{minipage}

\paragraph{Parallelization}

\pysr is parallelized by populations of expressions which evolve independently, which was described as the basis of allowing large-scale parallelization in~\cite{goldbergComparativeAnalysisSelection1991}.
In this framework, a population of expressions is dispatched to a worker (either a thread, if single-node, or a process, if multi-node), where it then evolves over many iterations.
Following a batch of iterations, this population is returned to the head worker, which performs several tasks:
(1) record the best expressions, at each complexity, found in the population as part of a global ``hall of fame;''
(2) record the current expressions so they can ``migrate'' to other populations; and, finally,
(3) randomly ``migrate'' expressions from other populations, and the global hall of fame, to this population.
After these steps are finished, this population is once again dispatched to a worker where it undergoes further evolution.
This parallelization is asynchronous, and different populations may complete evolution at different speeds without slowing down others.
The technique is largely similar to the method described in~\cite{schmidtDistillingFreeFormNatural2009}, although here we have modified the migration step to also migrate based on a global, permanent hall of fame, rather than exclusively migrating current population members.
Reintroducing hall of fame members can significantly speed up the search.

\subsection{Interfaces}
\label{sec:interfaces}

\pysr has export functionality to several popular Python libraries and other formats.
\pysr by default will export any discovered expression to SymPy~\cite{sympy}, which, by extension, can convert expressions to LaTeX for including in research reports.
The mechanism for this functionality is fairly simple: using a string representation of the recovered expression, \pysr replaces symbol names with the SymPy equivalents (e.g., \texttt{cos} would become \texttt{sympy.cos}),
and evaluates it with the Python interpreter.
As a caveat of this, the user must define any custom operator using SymPy functions, in addition to Julia functions.

A similar technique---evaluating the string representation for an expression---is used for exporting expressions to a callable format in NumPy, PyTorch, and JAX~\cite{numpy,torch,jax}---each of which take vector input.
For NumPy, constants in an expression are embedded inside the callable function.
However, for PyTorch and JAX, since one may wish to re-optimize the constants in an expression (for example, as is done in~\cite{lemosRediscoveringNewtonGravity2022}), the constants are trainable.
In PyTorch, this amounts to creating a PyTorch \texttt{nn.Parameter} for the constants, which causes PyTorch to track gradients with respect to the parameters, so they can be trained via gradient descent.
Since JAX is a functional language, the parameters are instead exported as a vector, which the user then will pass into the callable function.
This is similar to how deep learning frameworks in JAX are trained.

\subsection{Custom Operators}
\label{sec:custom_operators}

The occurrence rate of a particular mathematical operator in a set of models varies by the scientific field those models describe.
For example, models in Epidemiology commonly use exponentials, indicating exponential growth or decay of a contagious disease, but may rarely use a Bessel function — which would be more common in in classical physics applications.
The entire reason that SR produces interpretable models is that the generated expressions make use of operators which are common in a particular field, and a domain scientist may be able to see connections with existing models.
In some ways, this process relates to language: black box machine learning models represent functions in a language uninterpretable to a scientist, whereas SR uses the language in place.
Furthermore, due to the modular nature of scientific modelling---many complex models are built on-top of existing simple models for particular subsystems---it also makes sense from a standpoint of improving model accuracy to make use of common operators (for an interesting explicit example of this, see~\cite{guimeraBayesianMachineScientist2020}).

\pysr does not, per se, have any built-in operators.
Due to the just-in-time compiled nature of Julia, any real scalar function from the entirety of the Julia Base language is available, and can be compiled into the search code at runtime.
This includes commonly-used operators such as
\texttt{+},
\texttt{-},
\texttt{*},
\texttt{/},
$\wedge$,
\texttt{exp},
\texttt{log},
\texttt{sin},
\texttt{cos},
\texttt{tan},
\texttt{abs},
and many others.

Furthermore, since many domains of science have operators that are unique to their field, it is possible to pass an arbitrary Julia function as an operator, whether it be a binary operator (with two arguments) or a unary operator (with one argument).
Any function of the form
$f:\mathbb{R}\rightarrow \mathbb{R}$ or $\mathbb{R}^2\rightarrow\mathbb{R}$, whether
continuous or not,
can be used as a user-defined operator.
The function need only be designed for scalar input and output, and PySR will use Julia to automatically compile and vectorize it, generating SIMD (Single Instruction, Multiple Data) instructions when possible.
In \pysr, an example of this would be:
\begin{minipage}{0.95\linewidth}
\begin{minted}{python}
op = "special(x, y) = cos(x) * (x + y)"
model = PySRRegressor(binary_operators=[op])
\end{minted}
\end{minipage}
where the string \texttt{special(x, y) = cos(x) * (x + y)} is Julia code giving a function definition.
This would define a binary operator \texttt{special} that would be compiled into the search code.
To enable custom operators to be defined in the various export functionality, the user must also define equivalent operators in SymPy (here, \texttt{lambda x, y: sympy.cos(x) * (x + y)}), as well as JAX or PyTorch versions if the user wishes to export to those frameworks as well.

As an example in science, it is very common in astrophysics to see ``broken power laws'': power laws whose exponent takes on different values in the parameter space.
This could be defined in \pysr by enabling the power law operator $\wedge$, and then giving the string \texttt{cond(x, y) = x < 0 ? 0 : y}, which defines a conditional branch of an expression given some expression defined in the \texttt{x} variable (using the ternary operator \texttt{condition ? value1 : value2}).
For example, \texttt{pow(x, cond(x - 5, 3.4) - 2.1)} would define the broken power law:
$$
\left\{
\begin{array}{cr}
x \geq 5, & x^{1.3}\\
x < 5, & x^{4.5}.
\end{array}
\right.
$$
The importance of defining custom operators is that there is no standard set of operators which the library is specifically tuned for; any operator common in a particular field is feasible to implement (so long as it can take 1-2 scalar arguments).

\subsection{Custom Losses}
\label{sec:custom_losses}

In many machine learning toy datasets for benchmarking regression algorithms, Mean-Square-Error (MSE or $L_2$ loss) is typically used as a learning objective~\cite{grinsztajnWhyTreebasedModels2022}.
In a Bayesian framework, MSE is equivalent to assuming every data point is Gaussian distributed, with equal variance per point.
Minimizing MSE is equivalent to maximizing the Gaussian log-likelihood.
However, in science, one typically works with 
a likelihood that is very specific to a particular problem,
and this is often non-Gaussian.
Therefore, it is important for an SR package to allow for custom loss functions.
\pysr implements this in a way that is very similar to that of custom operators (see \cref{sec:custom_operators}).
Given a string such as ``\texttt{loss(x, y) = abs(x - y)}'', \pysr will pass this to the Julia backend, which will automatically vectorize it and use it as a loss function throughout the search process.
In a Bayesian context, this would allow one to define arbitrary likelihoods, even for very complex branching logic.
This also works for weighted losses, such as ``\texttt{loss(x, y, w) = abs(x - y) * w}''

\section{Evaluation}
\label{sec:evaluation}

In \cref{sec:intro_pysr}, we discussed practical issues with discovering symbolic expressions in the sciences.
Here, we present a comparison of \pysr with existing tools in terms of addressing these concerns.

\subsection{High Level Comparison}

First, we present a high level comparison of the various SR tools available.
These qualitative points are made using the desirable features given in \cref{sec:intro_pysr}.

\newcommand\extranotes{
& Expressivity scores: (1a) Pre-trained on equations generated from limited prior.
(1b) Basis of fixed expressions, combined in a linear sum.
(2) Flexible basis of expressions, with variable internal coefficients.
(3) Any scalar tree, with binary and unary operators.
(4) Any scalar tree, with custom operators allowed.
(5) Any scalar tree, with n-ary operators.
(6) Scalar/vector/tensor expressions of any arity.
\\
$\ast$ & Note that the ``Symbolic Distillation'' method from~\cite{cranmerDiscoveringSymbolicModels2020} is not an algorithm itself; it can be applied to any SR technique. Applying this general method to a specific technique will inherit a {\color{green}$\checkmark$} from the Symbolic Distillation column, if given. However, in general, this technique is easiest with those methods which have deep learning export.\\
}

\onecolumn
\begin{landscape}
\begin{table*} \centering
    \scriptsize
  \begin{tabularx}{\linewidth}{ll|>{\hsize=0.1\hsize\linewidth=\hsize\centering\arraybackslash}X>{\hsize=0.1\hsize\linewidth=\hsize\centering\arraybackslash}X>{\hsize=0.1\hsize\linewidth=\hsize\centering\arraybackslash}X>{\hsize=0.1\hsize\linewidth=\hsize\centering\arraybackslash}X>{\hsize=0.1\hsize\linewidth=\hsize\centering\arraybackslash}X>{\hsize=0.1\hsize\linewidth=\hsize\centering\arraybackslash}X>{\hsize=0.1\hsize\linewidth=\hsize\centering\arraybackslash}X>{\hsize=0.1\hsize\linewidth=\hsize\centering\arraybackslash}X>{\hsize=0.1\hsize\linewidth=\hsize\centering\arraybackslash}X>{\hsize=0.1\hsize\linewidth=\hsize\centering\arraybackslash}X>{\hsize=0.1\hsize\linewidth=\hsize\centering\arraybackslash}X|>{\hsize=0.1\hsize\linewidth=\hsize\centering\arraybackslash}X}
\rowcolor{gray!50}
\toprule
 &  & \rotatebox[origin=c]{90}{\ \textbf{PySR}\ } & \rotatebox[origin=c]{90}{\ Eureqa\ } & \rotatebox[origin=c]{90}{\ GPLearn\ } & \rotatebox[origin=c]{90}{\ AI Feynman\ } & \rotatebox[origin=c]{90}{\ Operon\ } & \rotatebox[origin=c]{90}{\ DSR\ } & \rotatebox[origin=c]{90}{\ PySINDy\ } & \rotatebox[origin=c]{90}{\ EQL\ } & \rotatebox[origin=c]{90}{\ QLattice\ } & \rotatebox[origin=c]{90}{\ SR-Transformer\ } & \rotatebox[origin=c]{90}{\ GP-GOMEA\ } & \rotatebox[origin=c]{90}{\ \begin{tabular}{@{}c@{}}Symbolic \\ Distillation$\ast$\end{tabular}\ }\\
\cellcolor[gray]{0.95}& \cellcolor[gray]{1.0}{Compiled} & \cellcolor[gray]{1.0}${\color{green}\checkmark}$ & \cellcolor[gray]{1.0}${\color{green}\checkmark}$ & \cellcolor[gray]{1.0}${\color{red}\cross}$ & \cellcolor[gray]{1.0}${\color{red}\cross}$ & \cellcolor[gray]{1.0}${\color{green}\checkmark}$ & \cellcolor[gray]{1.0}${\color{red}\cross}$ & \cellcolor[gray]{1.0}${\color{red}\cross}$ & \cellcolor[gray]{1.0}${\color{green}\checkmark}$ & \cellcolor[gray]{1.0}${\color{green}\checkmark}$ & \cellcolor[gray]{1.0}${\color{green}\checkmark}$ & \cellcolor[gray]{1.0}${\color{green}\checkmark}$ & \cellcolor[gray]{1.0}-\\
\cellcolor[gray]{0.95}& \cellcolor[gray]{0.8}{Multi-core} & \cellcolor[gray]{0.8}${\color{green}\checkmark}$ & \cellcolor[gray]{0.8}${\color{green}\checkmark}$ & \cellcolor[gray]{0.8}${\color{green}\checkmark}$ & \cellcolor[gray]{0.8}${\color{green}\checkmark}$ & \cellcolor[gray]{0.8}${\color{green}\checkmark}$ & \cellcolor[gray]{0.8}${\color{red}\cross}$ & \cellcolor[gray]{0.8}${\color{green}\checkmark}$ & \cellcolor[gray]{0.8}${\color{green}\checkmark}$ & \cellcolor[gray]{0.8}${\color{green}\checkmark}$ & \cellcolor[gray]{0.8}${\color{green}\checkmark}$ & \cellcolor[gray]{0.8}${\color{green}\checkmark}$ & \cellcolor[gray]{0.8}${\color{green}\checkmark}$\\
\cellcolor[gray]{0.95}& \cellcolor[gray]{1.0}{Multi-node} & \cellcolor[gray]{1.0}${\color{green}\checkmark}$ & \cellcolor[gray]{1.0}${\color{red}\cross}$ & \cellcolor[gray]{1.0}${\color{red}\cross}$ & \cellcolor[gray]{1.0}${\color{red}\cross}$ & \cellcolor[gray]{1.0}${\color{red}\cross}$ & \cellcolor[gray]{1.0}${\color{red}\cross}$ & \cellcolor[gray]{1.0}${\color{red}\cross}$ & \cellcolor[gray]{1.0}${\color{red}\cross}$ & \cellcolor[gray]{1.0}${\color{red}\cross}$ & \cellcolor[gray]{1.0}${\color{red}\cross}$ & \cellcolor[gray]{1.0}${\color{red}\cross}$ & \cellcolor[gray]{1.0}-\\
\multirow{-4}{*}{\cellcolor[gray]{0.95}\textbf{Scalability}} & \cellcolor[gray]{0.8}{GPU-capable} & \cellcolor[gray]{0.8}${\color{red}\cross}$ & \cellcolor[gray]{0.8}${\color{red}\cross}$ & \cellcolor[gray]{0.8}${\color{red}\cross}$ & \cellcolor[gray]{0.8}$\ast$I & \cellcolor[gray]{0.8}${\color{red}\cross}$ & \cellcolor[gray]{0.8}${\color{red}\cross}$ & \cellcolor[gray]{0.8}${\color{green}\checkmark}$ & \cellcolor[gray]{0.8}${\color{green}\checkmark}$ & \cellcolor[gray]{0.8}${\color{red}\cross}$ & \cellcolor[gray]{0.8}${\color{green}\checkmark}$ & \cellcolor[gray]{0.8}${\color{red}\cross}$ & \cellcolor[gray]{0.8}${\color{green}\checkmark}$\\
\cellcolor[gray]{0.85}& \cellcolor[gray]{1.0}{No pre-training} & \cellcolor[gray]{1.0}${\color{green}\checkmark}$ & \cellcolor[gray]{1.0}${\color{green}\checkmark}$ & \cellcolor[gray]{1.0}${\color{green}\checkmark}$ & \cellcolor[gray]{1.0}${\color{green}\checkmark}$ & \cellcolor[gray]{1.0}${\color{green}\checkmark}$ & \cellcolor[gray]{1.0}${\color{green}\checkmark}$ & \cellcolor[gray]{1.0}${\color{green}\checkmark}$ & \cellcolor[gray]{1.0}${\color{green}\checkmark}$ & \cellcolor[gray]{1.0}${\color{green}\checkmark}$ & \cellcolor[gray]{1.0}${\color{red}\cross}$ & \cellcolor[gray]{1.0}${\color{green}\checkmark}$ & \cellcolor[gray]{1.0}-\\
\cellcolor[gray]{0.85}& \cellcolor[gray]{0.8}{Denoising} & \cellcolor[gray]{0.8}${\color{green}\checkmark}$ & \cellcolor[gray]{0.8}${\color{green}\checkmark}$ & \cellcolor[gray]{0.8}${\color{red}\cross}$ & \cellcolor[gray]{0.8}${\color{red}\cross}$ & \cellcolor[gray]{0.8}${\color{red}\cross}$ & \cellcolor[gray]{0.8}${\color{red}\cross}$ & \cellcolor[gray]{0.8}$\ast$II & \cellcolor[gray]{0.8}${\color{red}\cross}$ & \cellcolor[gray]{0.8}? & \cellcolor[gray]{0.8}${\color{red}\cross}$ & \cellcolor[gray]{0.8}${\color{red}\cross}$ & \cellcolor[gray]{0.8}${\color{green}\checkmark}$\\
\cellcolor[gray]{0.85}& \cellcolor[gray]{1.0}{Feature selection} & \cellcolor[gray]{1.0}${\color{green}\checkmark}$ & \cellcolor[gray]{1.0}${\color{green}\checkmark}$ & \cellcolor[gray]{1.0}${\color{red}\cross}$ & \cellcolor[gray]{1.0}${\color{green}\checkmark}$ & \cellcolor[gray]{1.0}${\color{red}\cross}$ & \cellcolor[gray]{1.0}${\color{green}\checkmark}$ & \cellcolor[gray]{1.0}$\ast$II & \cellcolor[gray]{1.0}${\color{red}\cross}$ & \cellcolor[gray]{1.0}${\color{green}\checkmark}$ & \cellcolor[gray]{1.0}${\color{red}\cross}$ & \cellcolor[gray]{1.0}${\color{red}\cross}$ & \cellcolor[gray]{1.0}${\color{green}\checkmark}$\\
\cellcolor[gray]{0.85}& \cellcolor[gray]{0.8}{Differential equations} & \cellcolor[gray]{0.8}${\color{red}\cross}$ & \cellcolor[gray]{0.8}${\color{green}\checkmark}$ & \cellcolor[gray]{0.8}${\color{red}\cross}$ & \cellcolor[gray]{0.8}${\color{red}\cross}$ & \cellcolor[gray]{0.8}${\color{red}\cross}$ & \cellcolor[gray]{0.8}${\color{red}\cross}$ & \cellcolor[gray]{0.8}${\color{green}\checkmark}$ & \cellcolor[gray]{0.8}${\color{green}\checkmark}$ & \cellcolor[gray]{0.8}${\color{red}\cross}$ & \cellcolor[gray]{0.8}${\color{red}\cross}$ & \cellcolor[gray]{0.8}${\color{red}\cross}$ & \cellcolor[gray]{0.8}${\color{green}\checkmark}$\\
\cellcolor[gray]{0.85}& \cellcolor[gray]{1.0}{High-dimensional} & \cellcolor[gray]{1.0}${\color{red}\cross}$ & \cellcolor[gray]{1.0}${\color{red}\cross}$ & \cellcolor[gray]{1.0}${\color{red}\cross}$ & \cellcolor[gray]{1.0}${\color{red}\cross}$ & \cellcolor[gray]{1.0}${\color{red}\cross}$ & \cellcolor[gray]{1.0}${\color{red}\cross}$ & \cellcolor[gray]{1.0}${\color{green}\checkmark}$ & \cellcolor[gray]{1.0}${\color{green}\checkmark}$ & \cellcolor[gray]{1.0}${\color{red}\cross}$ & \cellcolor[gray]{1.0}${\color{red}\cross}$ & \cellcolor[gray]{1.0}${\color{red}\cross}$ & \cellcolor[gray]{1.0}${\color{green}\checkmark}$\\
\multirow{-6}{*}{\cellcolor[gray]{0.85}\textbf{Practicality}} & \cellcolor[gray]{0.8}{Full Pareto curve} & \cellcolor[gray]{0.8}${\color{green}\checkmark}$ & \cellcolor[gray]{0.8}${\color{green}\checkmark}$ & \cellcolor[gray]{0.8}${\color{green}\checkmark}$ & \cellcolor[gray]{0.8}${\color{red}\cross}$ & \cellcolor[gray]{0.8}${\color{green}\checkmark}$ & \cellcolor[gray]{0.8}${\color{green}\checkmark}$ & \cellcolor[gray]{0.8}$\ast$II & \cellcolor[gray]{0.8}${\color{red}\cross}$ & \cellcolor[gray]{0.8}${\color{green}\checkmark}$ & \cellcolor[gray]{0.8}${\color{red}\cross}$ & \cellcolor[gray]{0.8}${\color{green}\checkmark}$ & \cellcolor[gray]{0.8}${\color{red}\cross}$\\
\cellcolor[gray]{0.95}& \cellcolor[gray]{1.0}{API} & \cellcolor[gray]{1.0}${\color{green}\checkmark}$ & \cellcolor[gray]{1.0}${\color{red}\cross}$ & \cellcolor[gray]{1.0}${\color{green}\checkmark}$ & \cellcolor[gray]{1.0}${\color{red}\cross}$ & \cellcolor[gray]{1.0}${\color{green}\checkmark}$ & \cellcolor[gray]{1.0}${\color{green}\checkmark}$ & \cellcolor[gray]{1.0}${\color{green}\checkmark}$ & \cellcolor[gray]{1.0}${\color{green}\checkmark}$ & \cellcolor[gray]{1.0}${\color{green}\checkmark}$ & \cellcolor[gray]{1.0}${\color{green}\checkmark}$ & \cellcolor[gray]{1.0}${\color{green}\checkmark}$ & \cellcolor[gray]{1.0}-\\
\cellcolor[gray]{0.95}& \cellcolor[gray]{0.8}{SymPy Interface} & \cellcolor[gray]{0.8}${\color{green}\checkmark}$ & \cellcolor[gray]{0.8}${\color{red}\cross}$ & \cellcolor[gray]{0.8}${\color{red}\cross}$ & \cellcolor[gray]{0.8}${\color{green}\checkmark}$ & \cellcolor[gray]{0.8}${\color{green}\checkmark}$ & \cellcolor[gray]{0.8}${\color{red}\cross}$ & \cellcolor[gray]{0.8}${\color{red}\cross}$ & \cellcolor[gray]{0.8}${\color{red}\cross}$ & \cellcolor[gray]{0.8}${\color{green}\checkmark}$ & \cellcolor[gray]{0.8}${\color{green}\checkmark}$ & \cellcolor[gray]{0.8}${\color{green}\checkmark}$ & \cellcolor[gray]{0.8}-\\
\multirow{-3}{*}{\cellcolor[gray]{0.95}\textbf{Interfacing}} & \cellcolor[gray]{1.0}{Deep Learning export} & \cellcolor[gray]{1.0}${\color{green}\checkmark}$ & \cellcolor[gray]{1.0}${\color{red}\cross}$ & \cellcolor[gray]{1.0}${\color{red}\cross}$ & \cellcolor[gray]{1.0}${\color{red}\cross}$ & \cellcolor[gray]{1.0}${\color{red}\cross}$ & \cellcolor[gray]{1.0}${\color{red}\cross}$ & \cellcolor[gray]{1.0}${\color{red}\cross}$ & \cellcolor[gray]{1.0}$\ast$III & \cellcolor[gray]{1.0}${\color{red}\cross}$ & \cellcolor[gray]{1.0}$\ast$III & \cellcolor[gray]{1.0}${\color{red}\cross}$ & \cellcolor[gray]{1.0}-\\
\cellcolor[gray]{0.85}& \cellcolor[gray]{0.8}{Expressivity score} & \cellcolor[gray]{0.8}4 & \cellcolor[gray]{0.8}5 & \cellcolor[gray]{0.8}4 & \cellcolor[gray]{0.8}3 & \cellcolor[gray]{0.8}3 & \cellcolor[gray]{0.8}3 & \cellcolor[gray]{0.8}1b & \cellcolor[gray]{0.8}2 & \cellcolor[gray]{0.8}3 & \cellcolor[gray]{0.8}1a & \cellcolor[gray]{0.8}3 & \cellcolor[gray]{0.8}6\\
\cellcolor[gray]{0.85}& \cellcolor[gray]{1.0}{Open-source} & \cellcolor[gray]{1.0}${\color{green}\checkmark}$ & \cellcolor[gray]{1.0}${\color{red}\cross}$ & \cellcolor[gray]{1.0}${\color{green}\checkmark}$ & \cellcolor[gray]{1.0}${\color{green}\checkmark}$ & \cellcolor[gray]{1.0}${\color{green}\checkmark}$ & \cellcolor[gray]{1.0}${\color{green}\checkmark}$ & \cellcolor[gray]{1.0}${\color{green}\checkmark}$ & \cellcolor[gray]{1.0}${\color{green}\checkmark}$ & \cellcolor[gray]{1.0}${\color{red}\cross}$ & \cellcolor[gray]{1.0}${\color{green}\checkmark}$ & \cellcolor[gray]{1.0}${\color{green}\checkmark}$ & \cellcolor[gray]{1.0}${\color{green}\checkmark}$\\
\cellcolor[gray]{0.85}& \cellcolor[gray]{0.8}{Real Constants} & \cellcolor[gray]{0.8}${\color{green}\checkmark}$ & \cellcolor[gray]{0.8}${\color{green}\checkmark}$ & \cellcolor[gray]{0.8}${\color{green}\checkmark}$ & \cellcolor[gray]{0.8}${\color{red}\cross}$ & \cellcolor[gray]{0.8}${\color{green}\checkmark}$ & \cellcolor[gray]{0.8}${\color{green}\checkmark}$ & \cellcolor[gray]{0.8}$\ast$II & \cellcolor[gray]{0.8}${\color{green}\checkmark}$ & \cellcolor[gray]{0.8}${\color{green}\checkmark}$ & \cellcolor[gray]{0.8}${\color{green}\checkmark}$ & \cellcolor[gray]{0.8}${\color{green}\checkmark}$ & \cellcolor[gray]{0.8}-\\
\cellcolor[gray]{0.85}& \cellcolor[gray]{1.0}{Custom operators} & \cellcolor[gray]{1.0}${\color{green}\checkmark}$ & \cellcolor[gray]{1.0}${\color{red}\cross}$ & \cellcolor[gray]{1.0}${\color{green}\checkmark}$ & \cellcolor[gray]{1.0}${\color{red}\cross}$ & \cellcolor[gray]{1.0}${\color{red}\cross}$ & \cellcolor[gray]{1.0}${\color{red}\cross}$ & \cellcolor[gray]{1.0}$\ast$II & \cellcolor[gray]{1.0}${\color{red}\cross}$ & \cellcolor[gray]{1.0}${\color{red}\cross}$ & \cellcolor[gray]{1.0}${\color{red}\cross}$ & \cellcolor[gray]{1.0}${\color{red}\cross}$ & \cellcolor[gray]{1.0}-\\
\cellcolor[gray]{0.85}& \cellcolor[gray]{0.8}{Discontinuous operators} & \cellcolor[gray]{0.8}${\color{green}\checkmark}$ & \cellcolor[gray]{0.8}${\color{green}\checkmark}$ & \cellcolor[gray]{0.8}${\color{green}\checkmark}$ & \cellcolor[gray]{0.8}${\color{red}\cross}$ & \cellcolor[gray]{0.8}${\color{red}\cross}$ & \cellcolor[gray]{0.8}${\color{red}\cross}$ & \cellcolor[gray]{0.8}$\ast$II & \cellcolor[gray]{0.8}${\color{red}\cross}$ & \cellcolor[gray]{0.8}${\color{red}\cross}$ & \cellcolor[gray]{0.8}${\color{red}\cross}$ & \cellcolor[gray]{0.8}${\color{red}\cross}$ & \cellcolor[gray]{0.8}-\\
\cellcolor[gray]{0.85}& \cellcolor[gray]{1.0}{Custom losses} & \cellcolor[gray]{1.0}${\color{green}\checkmark}$ & \cellcolor[gray]{1.0}${\color{green}\checkmark}$ & \cellcolor[gray]{1.0}${\color{green}\checkmark}$ & \cellcolor[gray]{1.0}${\color{red}\cross}$ & \cellcolor[gray]{1.0}${\color{red}\cross}$ & \cellcolor[gray]{1.0}${\color{green}\checkmark}$ & \cellcolor[gray]{1.0}${\color{red}\cross}$ & \cellcolor[gray]{1.0}${\color{red}\cross}$ & \cellcolor[gray]{1.0}${\color{red}\cross}$ & \cellcolor[gray]{1.0}${\color{red}\cross}$ & \cellcolor[gray]{1.0}${\color{red}\cross}$ & \cellcolor[gray]{1.0}${\color{green}\checkmark}$\\
\cellcolor[gray]{0.85}& \cellcolor[gray]{0.8}{Symbolic Constraints} & \cellcolor[gray]{0.8}${\color{green}\checkmark}$ & \cellcolor[gray]{0.8}${\color{red}\cross}$ & \cellcolor[gray]{0.8}${\color{green}\checkmark}$ & \cellcolor[gray]{0.8}${\color{red}\cross}$ & \cellcolor[gray]{0.8}${\color{red}\cross}$ & \cellcolor[gray]{0.8}${\color{green}\checkmark}$ & \cellcolor[gray]{0.8}${\color{red}\cross}$ & \cellcolor[gray]{0.8}${\color{red}\cross}$ & \cellcolor[gray]{0.8}${\color{red}\cross}$ & \cellcolor[gray]{0.8}${\color{red}\cross}$ & \cellcolor[gray]{0.8}${\color{red}\cross}$ & \cellcolor[gray]{0.8}${\color{green}\checkmark}$\\
\cellcolor[gray]{0.85}& \cellcolor[gray]{1.0}{Custom complexity} & \cellcolor[gray]{1.0}${\color{green}\checkmark}$ & \cellcolor[gray]{1.0}${\color{green}\checkmark}$ & \cellcolor[gray]{1.0}${\color{green}\checkmark}$ & \cellcolor[gray]{1.0}${\color{red}\cross}$ & \cellcolor[gray]{1.0}${\color{red}\cross}$ & \cellcolor[gray]{1.0}${\color{red}\cross}$ & \cellcolor[gray]{1.0}${\color{red}\cross}$ & \cellcolor[gray]{1.0}${\color{red}\cross}$ & \cellcolor[gray]{1.0}${\color{red}\cross}$ & \cellcolor[gray]{1.0}${\color{red}\cross}$ & \cellcolor[gray]{1.0}${\color{red}\cross}$ & \cellcolor[gray]{1.0}-\\
\multirow{-9}{*}{\cellcolor[gray]{0.85}\textbf{Extensibility}} & \cellcolor[gray]{0.8}{Custom types} & \cellcolor[gray]{0.8}${\color{green}\checkmark}$ & \cellcolor[gray]{0.8}${\color{red}\cross}$ & \cellcolor[gray]{0.8}${\color{red}\cross}$ & \cellcolor[gray]{0.8}${\color{red}\cross}$ & \cellcolor[gray]{0.8}${\color{red}\cross}$ & \cellcolor[gray]{0.8}${\color{red}\cross}$ & \cellcolor[gray]{0.8}${\color{red}\cross}$ & \cellcolor[gray]{0.8}${\color{red}\cross}$ & \cellcolor[gray]{0.8}${\color{red}\cross}$ & \cellcolor[gray]{0.8}${\color{red}\cross}$ & \cellcolor[gray]{0.8}${\color{red}\cross}$ & \cellcolor[gray]{0.8}${\color{red}\cross}$\\
\cellcolor[gray]{0.95}& \cellcolor[gray]{1.0}{Citation} & \cellcolor[gray]{1.0}[self] & \cellcolor[gray]{1.0}\cite{schmidtDistillingFreeFormNatural2009} & \cellcolor[gray]{1.0}- & \cellcolor[gray]{1.0}\cite{udrescuAIFeynmanPhysicsinspired2020} & \cellcolor[gray]{1.0}\cite{burlacuOperonEfficientGenetic2020} & \cellcolor[gray]{1.0}\cite{petersenDeepSymbolicRegression2021} & \cellcolor[gray]{1.0}\cite{kaptanogluPySINDyComprehensivePython2021} & \cellcolor[gray]{1.0}\cite{sahooLearningEquationsExtrapolation2018} & \cellcolor[gray]{1.0}\cite{brolosApproachSymbolicRegression2021} & \cellcolor[gray]{1.0}\cite{kamiennyEndtoendSymbolicRegression2022} & \cellcolor[gray]{1.0}\cite{virgolinImprovingModelbasedGenetic2021} & \cellcolor[gray]{1.0}\cite{cranmerDiscoveringSymbolicModels2020}\\
\multirow{-2}{*}{\cellcolor[gray]{0.95}\textbf{-}} & \cellcolor[gray]{0.8}{Code} & \cellcolor[gray]{0.8}\href{https://github.com/MilesCranmer/PySR}{\faUnlock} & \cellcolor[gray]{0.8}\faLock & \cellcolor[gray]{0.8}\href{https://github.com/trevorstephens/gplearn}{\faUnlock} & \cellcolor[gray]{0.8}\href{https://github.com/SJ001/AI-Feynman}{\faUnlock} & \cellcolor[gray]{0.8}\href{https://github.com/heal-research/operon}{\faUnlock} & \cellcolor[gray]{0.8}\href{https://github.com/brendenpetersen/deep-symbolic-optimization}{\faUnlock} & \cellcolor[gray]{0.8}\href{https://github.com/dynamicslab/pysindy}{\faUnlock} & \cellcolor[gray]{0.8}\href{https://github.com/martius-lab/EQL}{\faUnlock} & \cellcolor[gray]{0.8}\faLock & \cellcolor[gray]{0.8}\href{https://github.com/pakamienny/e2e_transformer}{\faUnlock} & \cellcolor[gray]{0.8}\href{https://github.com/marcovirgolin/GP-GOMEA}{\faUnlock} & \cellcolor[gray]{0.8}\href{https://github.com/MilesCranmer/symbolic_deep_learning}{\faUnlock}\\
\bottomrule
\end{tabularx}
\begin{tabularx}{\linewidth}{lX}
\toprule
\extranotes
$\ast$I & Only the symmetry discovery module is GPU-capable.\\
$\ast$II & Conceptually different, as is a linear basis of static nonlinear expressions.\\
$\ast$III & Is itself a neural network.\\
\bottomrule
\end{tabularx}
\unskip\label{output/comparison_table.tex}\unskip%

    \caption{Note that many instances of {\color{red}$\cross$} are \textbf{purely software limitations}. For example, most non-compiled algorithms \textit{could} support custom losses and operators, but few make this easily configurable via an API, which is important for practical use in science. Open source code can be found by clicking on each \href{https://github.com/MilesCranmer/PySR}{\faUnlock\xspace} icon.
    }
    \label{tbl:comparison_table}
\end{table*}
\end{landscape}
\twocolumn

\newcommand\fullbench{Empirical Science Symbolic Regression Benchmark\xspace}
\newcommand\bench{EmpiricalBench\xspace}

\subsection{\bench: \fullbench}
\label{sec:new_benchmark}

Existing SR benchmarks have certain simplifications compared to the datasets used to historically discover well-known empirical equations.
When discovering a new expression, one does not actually know the physical constants in the expression, and one must have to learn real constants.
For example, in the Feynman benchmark dataset of~\cite{udrescuAIFeynmanPhysicsinspired2020}, all expressions are listed with the associated physical constants---but these constants originally had to be discovered along with the equation.
Similarly, in the commonly-used synthetic ``Nguyen'' benchmark~\cite{nguyenquanguySemanticallybasedCrossoverGenetic2011}, which includes expressions such as $F_9=\sin(x)+\sin(y^2)$, there are no non-integral constants.
The ``SRBench'' competition of~\cite{defrancaInterpretableSymbolicRegression2023} was meticulous in its comparisons of different methods, but only contained a single real-world task (with an unknown ground truth), with all other expressions being synthetic.
This benchmark improved on others in that it included tunable amounts of noise.
However, the noise models are very simple: perfectly Gaussian and including no heteroscedasticity.
Although such synthetic benchmarks have their uses in SR research, these types of benchmarks are often not entirely indicative of the challenges faced in real-world scientific discovery.
Therefore, here we introduce a new benchmark which attempts to accurately portray the empirical equation discovery step of science.
Every equation in this dataset is a real empirical expression that a scientist has at one point discovered from experimental, noisy, and imperfect data.

\begin{figure*}[h!]
    \script{equations.py}
    \begin{center}
        \includegraphics[width=0.95\linewidth]{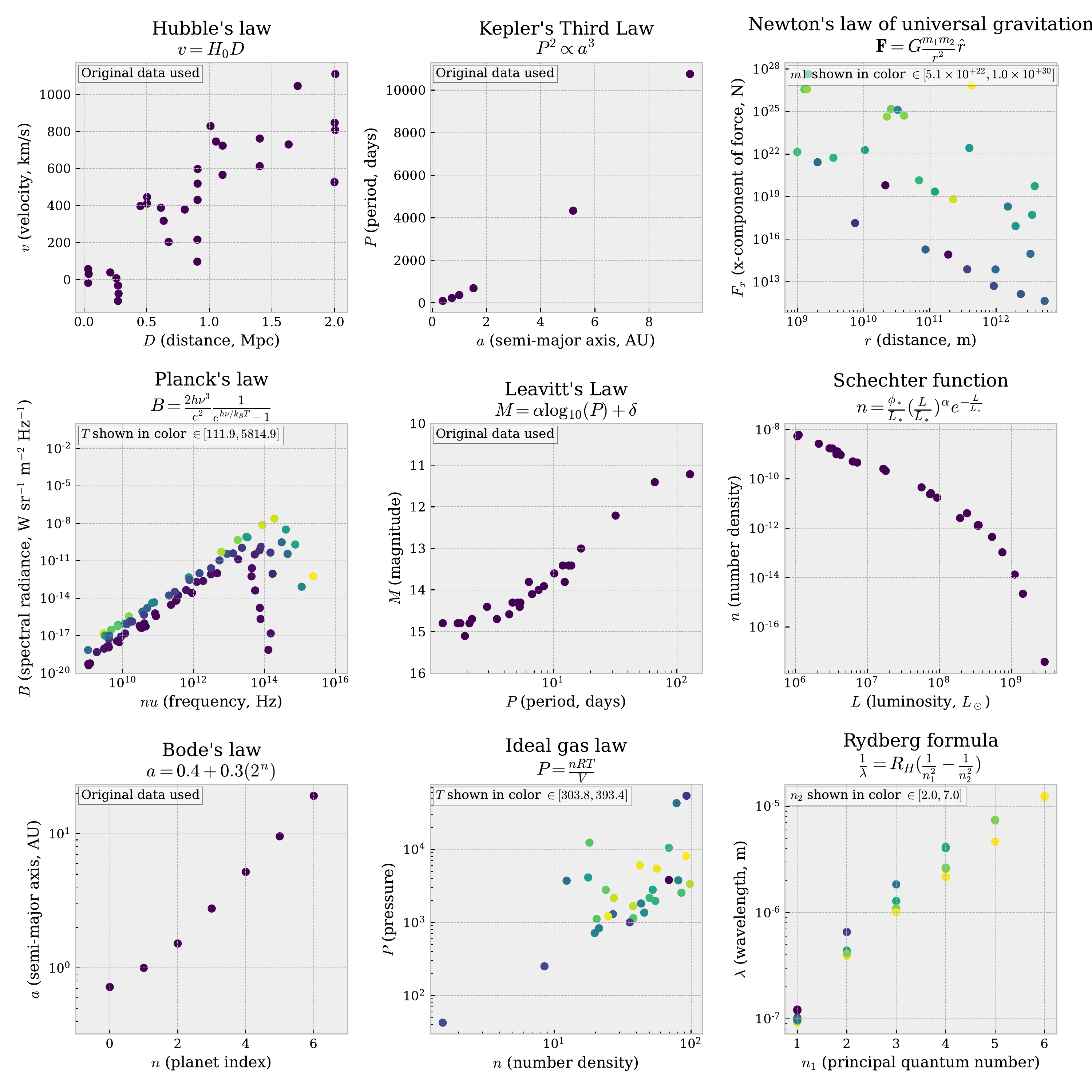}
    \end{center}
    \caption{Visualization of all the data in \bench, an SR benchmark for science.
    Color is used to denote additional variables in the cases of relations which depend on more than two inputs.
    Original data in the discovery of each law is used where easily available.
    Otherwise, data is generated from the formula with realistic ranges of variables, with a level of noise applied.
    }
    \label{fig:bench_data}
\end{figure*}

If an algorithm is to be considered to be useful for equation discovery in the sciences, it should be able to discover such relations.
Data for this benchmark is shown in \cref{fig:bench_data}, with the equations and
citations given in \cref{tbl:bench_data_table}.
Where original data was used, it was either: (a) taken from an original dataset available publicly, or (b) digitized from a table or plot using \texttt{WebPlotDigitizer}~\cite{rohatgiWebplotdigitizerVersion2022}.
For example, the data for Hubble's Law was manually extracted from the original 1929 paper~\citep[][]{hubbleRelationDistanceRadial1929}
Where original data was not available, data was generated from the equation with realistic ranges of variables, with noise applied.
The code and datasets used to generate this benchmark and evaluate are available in the paper repository, and are based on a fork of the SRBench competition~\cite{lacavaContemporarySymbolicRegression2021,defrancaInterpretableSymbolicRegression2023}.

\begin{table*}[h!]
    \centering
    \begin{tabular}{@{}lcc@{}}
        \toprule
        Name & Law & Early Citation \\
        \midrule
  Hubble's law & $v = H_0 D$ & \cite{hubbleRelationDistanceRadial1929}\\
Kepler's Third Law & $P^2 \propto a^3$ & \cite{keplerHarmonicesMundi1619}\\
Newton's law of universal gravitation & $\mathbf{F} = G \frac{m_1 m_2}{r^2} \hat{r}$ & \cite{newtonPhilosophiaeNaturalisPrincipia1687}\\
Planck's law & $B = \frac{2 h \nu^3}{c^2} \frac{1}{e^{h \nu/k_B T} - 1}$ & \cite{planckTheoryHeatRadiation1914}\\
Leavitt's Law & $M = \alpha\log_{10}(P) + \delta$ & \cite{leavittPeriods25Variable1912}\\
Schechter function & $n = \frac{\phi_\ast}{L_\ast} (\frac{L}{L_\ast})^{\alpha} e^{-\frac{L }{ L_\ast}}$ & \cite{pressFormationGalaxiesClusters1974}\\
Bode's law & $a = 0.4 + 0.3 (2^n)$ & \cite{bonnetContemplationNature1764}\\
Ideal gas law & $P = \frac{nRT}{V}$ & \cite{clapeyronMemoirePuissanceMotrice1835}\\
Rydberg formula & $\frac{1}{\lambda} = R_H(\frac{1}{n_1^2} - \frac{1}{n_2^2})$ & \cite{rydbergResearchesConstitutionSpectres1889}\unskip\label{output/table.tex}\unskip%
\\
        \bottomrule
    \end{tabular}
    \caption{Expressions in the \bench and associated with the datasets in \cref{fig:bench_data}. Each of these expressions was originally empirically discovered.}
    \label{tbl:bench_data_table}
\end{table*}

As noted in the ``Regression Target'' column, many of these laws which are easily expressed in logarithmic units are instead posed as a problem in observed units.
The search algorithm must discover this better unit by itself.

Many existing benchmarks for physical expression searches are based on theoretically-derived equations rather than empirically-observed and formulated. 
While these are likely useful for benchmarking some parts of the SR landscape, they do not target the aspect which SR would ultimately be used for: empirical discovery.
Thus, all the equations in this dataset were first empirically formulated before (and if) they were theoretically-derived.
Thus, the physical variables here directly correspond to observables.
Furthermore, this dataset does not include any relevant physical constants---the algorithm must find these automatically, as did the scientist who discovered the equation.

\subsubsection{Results on \bench}

We fork the SRBench competition repository~\cite{defrancaInterpretableSymbolicRegression2023}, as the authors have managed to perform the impressive task of aggregating a large variety of existing methods into a single repository with a common API.
However, here we aim to study the entire Pareto front of each algorithm, rather than only a single expression, as was done in SRBench.
By reviewing the discovered expressions over the entire Pareto front, we are not sensitive to the somewhat arbitrary choice of selection criteria, which differs by method. 
We can more accurately gauge performance of the algorithms themselves.
Thus, we modify the interface of every algorithm to return a Pareto front of expressions, or, in the cases where it was unavailable, instead we return a list of several candidate expressions.
We were able to successfully test the algorithms \pysr, Operon~\cite{burlacuOperonEfficientGenetic2020}, DSR~\cite{petersenDeepSymbolicRegression2021}, EQL~\cite{sahooLearningEquationsExtrapolation2018}, QLattice~\cite{brolosApproachSymbolicRegression2021}, and SR-Transformer~\cite{kamiennyEndtoendSymbolicRegression2022} on \bench.
Other codes were either not included in SRBench already, incompatible with our tests, or were otherwise unable to be configured on our system after significant effort.

We allow every method to use 8 cores on an AMD Rome CPU running Rocky Linux, and allow them to search for up to 1 hour.
These are similar constraints to the SRBench competition, which had authors individually tune their codes for such a setting, and thus we consider these settings most fair.

\begin{table*}[h!]
    \centering
  \begin{tabularx}{\textwidth}{l|>{\hsize=0.15\hsize\linewidth=\hsize\centering\arraybackslash}X>{\hsize=0.15\hsize\linewidth=\hsize\centering\arraybackslash}X>{\hsize=0.15\hsize\linewidth=\hsize\centering\arraybackslash}X>{\hsize=0.15\hsize\linewidth=\hsize\centering\arraybackslash}X>{\hsize=0.15\hsize\linewidth=\hsize\centering\arraybackslash}X>{\hsize=0.15\hsize\linewidth=\hsize\centering\arraybackslash}X}\\
\toprule & \rotatebox[origin=c]{90}{\ \textbf{PySR}\ } & \rotatebox[origin=c]{90}{\ Operon\ } & \rotatebox[origin=c]{90}{\ DSR\ } & \rotatebox[origin=c]{90}{\ EQL\ } & \rotatebox[origin=c]{90}{\ QLattice\ } & \rotatebox[origin=c]{90}{\ SR-Transformer\ }\\
\midrule { Hubble} & \begin{tabular}[x]{@{}c@{}} {$  \color[rgb]{0.1450980392156863,0.7450980392156863,0.34509803921568627} \Large  \nicefrac{\bm{5}}{\bm{5}} $}\\ {\color{gray} \tiny (5, 0, 0, 0)}\end{tabular} & \begin{tabular}[x]{@{}c@{}} {$  \color[rgb]{0.4823529411764706,0.09411764705882353,0.09411764705882353} \Large  \nicefrac{\bm{0}}{\bm{5}} $}\\ {\color{gray} \tiny (0, 5, 0, 0)}\end{tabular} & \begin{tabular}[x]{@{}c@{}} {$  \color[rgb]{0.6784313725490196,0.45647058823529413,0.11764705882352941} \Large  \nicefrac{\bm{1}}{\bm{5}} $}\\ {\color{gray} \tiny (1, 0, 4, 0)}\end{tabular} & \begin{tabular}[x]{@{}c@{}} {$  \color[rgb]{0.4823529411764706,0.09411764705882353,0.09411764705882353} \Large  \nicefrac{\bm{0}}{\bm{5}} $}\\ {\color{gray} \tiny (0, 0, 0, 5)}\end{tabular} & \begin{tabular}[x]{@{}c@{}} {$  \color[rgb]{0.4823529411764706,0.09411764705882353,0.09411764705882353} \Large  \nicefrac{\bm{0}}{\bm{5}} $}\\ {\color{gray} \tiny (0, 5, 0, 0)}\end{tabular} & \begin{tabular}[x]{@{}c@{}} {$  \color[rgb]{0.4823529411764706,0.09411764705882353,0.09411764705882353} \Large  \nicefrac{\bm{0}}{\bm{5}} $}\\ {\color{gray} \tiny (0, 0, 0, 5)}\end{tabular}\\
{ Kepler} & \begin{tabular}[x]{@{}c@{}} {$  \color[rgb]{0.1450980392156863,0.7450980392156863,0.34509803921568627} \Large  \nicefrac{\bm{5}}{\bm{5}} $}\\ {\color{gray} \tiny (5, 0, 0, 0)}\end{tabular} & \begin{tabular}[x]{@{}c@{}} {$  \color[rgb]{0.4823529411764706,0.09411764705882353,0.09411764705882353} \Large  \nicefrac{\bm{0}}{\bm{5}} $}\\ {\color{gray} \tiny (0, 5, 0, 0)}\end{tabular} & \begin{tabular}[x]{@{}c@{}} {$  \color[rgb]{0.47607843137254907,0.8470588235294118,0.26823529411764707} \Large  \nicefrac{\bm{4}}{\bm{5}} $}\\ {\color{gray} \tiny (4, 1, 0, 0)}\end{tabular} & \begin{tabular}[x]{@{}c@{}} {$  \color[rgb]{0.4823529411764706,0.09411764705882353,0.09411764705882353} \Large  \nicefrac{\bm{0}}{\bm{5}} $}\\ {\color{gray} \tiny (0, 0, 2, 3)}\end{tabular} & \begin{tabular}[x]{@{}c@{}} {$  \color[rgb]{0.4823529411764706,0.09411764705882353,0.09411764705882353} \Large  \nicefrac{\bm{0}}{\bm{5}} $}\\ {\color{gray} \tiny (0, 0, 0, 5)}\end{tabular} & \begin{tabular}[x]{@{}c@{}} {$  \color[rgb]{0.4823529411764706,0.09411764705882353,0.09411764705882353} \Large  \nicefrac{\bm{0}}{\bm{5}} $}\\ {\color{gray} \tiny (0, 0, 0, 5)}\end{tabular}\\
{ Newton} & \begin{tabular}[x]{@{}c@{}} {$  \color[rgb]{0.1450980392156863,0.7450980392156863,0.34509803921568627} \Large  \nicefrac{\bm{5}}{\bm{5}} $}\\ {\color{gray} \tiny (5, 0, 0, 0)}\end{tabular} & \begin{tabular}[x]{@{}c@{}} {$  \color[rgb]{0.6784313725490196,0.45647058823529413,0.11764705882352941} \Large  \nicefrac{\bm{1}}{\bm{5}} $}\\ {\color{gray} \tiny (1, 2, 0, 2)}\end{tabular} & \begin{tabular}[x]{@{}c@{}} {$  \color[rgb]{0.6784313725490196,0.45647058823529413,0.11764705882352941} \Large  \nicefrac{\bm{1}}{\bm{5}} $}\\ {\color{gray} \tiny (1, 0, 4, 0)}\end{tabular} & \begin{tabular}[x]{@{}c@{}} {$  \color[rgb]{0.4823529411764706,0.09411764705882353,0.09411764705882353} \Large  \nicefrac{\bm{0}}{\bm{5}} $}\\ {\color{gray} \tiny (0, 0, 5, 0)}\end{tabular} & \begin{tabular}[x]{@{}c@{}} {$  \color[rgb]{0.4823529411764706,0.09411764705882353,0.09411764705882353} \Large  \nicefrac{\bm{0}}{\bm{5}} $}\\ {\color{gray} \tiny (0, 0, 0, 5)}\end{tabular} & \begin{tabular}[x]{@{}c@{}} {$  \color[rgb]{0.4823529411764706,0.09411764705882353,0.09411764705882353} \Large  \nicefrac{\bm{0}}{\bm{5}} $}\\ {\color{gray} \tiny (0, 0, 0, 5)}\end{tabular}\\
{ Planck} & \begin{tabular}[x]{@{}c@{}} {$  \color[rgb]{0.4823529411764706,0.09411764705882353,0.09411764705882353} \Large  \nicefrac{\bm{0}}{\bm{5}} $}\\ {\color{gray} \tiny (0, 0, 0, 5)}\end{tabular} & \begin{tabular}[x]{@{}c@{}} {$  \color[rgb]{0.4823529411764706,0.09411764705882353,0.09411764705882353} \Large  \nicefrac{\bm{0}}{\bm{5}} $}\\ {\color{gray} \tiny (0, 0, 0, 5)}\end{tabular} & \begin{tabular}[x]{@{}c@{}} {$  \color[rgb]{0.4823529411764706,0.09411764705882353,0.09411764705882353} \Large  \nicefrac{\bm{0}}{\bm{5}} $}\\ {\color{gray} \tiny (0, 0, 1, 4)}\end{tabular} & \begin{tabular}[x]{@{}c@{}} {$  \color[rgb]{0.4823529411764706,0.09411764705882353,0.09411764705882353} \Large  \nicefrac{\bm{0}}{\bm{5}} $}\\ {\color{gray} \tiny (0, 0, 5, 0)}\end{tabular} & \begin{tabular}[x]{@{}c@{}} {$  \color[rgb]{0.4823529411764706,0.09411764705882353,0.09411764705882353} \Large  \nicefrac{\bm{0}}{\bm{5}} $}\\ {\color{gray} \tiny (0, 0, 0, 5)}\end{tabular} & \begin{tabular}[x]{@{}c@{}} {$  \color[rgb]{0.4823529411764706,0.09411764705882353,0.09411764705882353} \Large  \nicefrac{\bm{0}}{\bm{5}} $}\\ {\color{gray} \tiny (0, 0, 0, 5)}\end{tabular}\\
{ Leavitt} & \begin{tabular}[x]{@{}c@{}} {$  \color[rgb]{0.1450980392156863,0.7450980392156863,0.34509803921568627} \Large  \nicefrac{\bm{5}}{\bm{5}} $}\\ {\color{gray} \tiny (5, 0, 0, 0)}\end{tabular} & \begin{tabular}[x]{@{}c@{}} {$  \color[rgb]{0.4823529411764706,0.09411764705882353,0.09411764705882353} \Large  \nicefrac{\bm{0}}{\bm{5}} $}\\ {\color{gray} \tiny (0, 0, 0, 5)}\end{tabular} & \begin{tabular}[x]{@{}c@{}} {$  \color[rgb]{0.1450980392156863,0.7450980392156863,0.34509803921568627} \Large  \nicefrac{\bm{5}}{\bm{5}} $}\\ {\color{gray} \tiny (5, 0, 0, 0)}\end{tabular} & \begin{tabular}[x]{@{}c@{}} {$  \color[rgb]{0.4823529411764706,0.09411764705882353,0.09411764705882353} \Large  \nicefrac{\bm{0}}{\bm{5}} $}\\ {\color{gray} \tiny (0, 0, 5, 0)}\end{tabular} & \begin{tabular}[x]{@{}c@{}} {$  \color[rgb]{0.4823529411764706,0.09411764705882353,0.09411764705882353} \Large  \nicefrac{\bm{0}}{\bm{5}} $}\\ {\color{gray} \tiny (0, 0, 0, 5)}\end{tabular} & \begin{tabular}[x]{@{}c@{}} {$  \color[rgb]{0.4823529411764706,0.09411764705882353,0.09411764705882353} \Large  \nicefrac{\bm{0}}{\bm{5}} $}\\ {\color{gray} \tiny (0, 0, 0, 5)}\end{tabular}\\
{ Schechter} & \begin{tabular}[x]{@{}c@{}} {$  \color[rgb]{0.1450980392156863,0.7450980392156863,0.34509803921568627} \Large  \nicefrac{\bm{5}}{\bm{5}} $}\\ {\color{gray} \tiny (5, 0, 0, 0)}\end{tabular} & \begin{tabular}[x]{@{}c@{}} {$  \color[rgb]{0.1450980392156863,0.7450980392156863,0.34509803921568627} \Large  \nicefrac{\bm{5}}{\bm{5}} $}\\ {\color{gray} \tiny (5, 0, 0, 0)}\end{tabular} & \begin{tabular}[x]{@{}c@{}} {$  \color[rgb]{0.1450980392156863,0.7450980392156863,0.34509803921568627} \Large  \nicefrac{\bm{5}}{\bm{5}} $}\\ {\color{gray} \tiny (5, 0, 0, 0)}\end{tabular} & \begin{tabular}[x]{@{}c@{}} {$  \color[rgb]{0.4823529411764706,0.09411764705882353,0.09411764705882353} \Large  \nicefrac{\bm{0}}{\bm{5}} $}\\ {\color{gray} \tiny (0, 0, 4, 1)}\end{tabular} & \begin{tabular}[x]{@{}c@{}} {$  \color[rgb]{0.1450980392156863,0.7450980392156863,0.34509803921568627} \Large  \nicefrac{\bm{5}}{\bm{5}} $}\\ {\color{gray} \tiny (5, 0, 0, 0)}\end{tabular} & \begin{tabular}[x]{@{}c@{}} {$  \color[rgb]{0.4823529411764706,0.09411764705882353,0.09411764705882353} \Large  \nicefrac{\bm{0}}{\bm{5}} $}\\ {\color{gray} \tiny (0, 0, 0, 5)}\end{tabular}\\
{ Bode} & \begin{tabular}[x]{@{}c@{}} {$  \color[rgb]{0.1450980392156863,0.7450980392156863,0.34509803921568627} \Large  \nicefrac{\bm{5}}{\bm{5}} $}\\ {\color{gray} \tiny (5, 0, 0, 0)}\end{tabular} & \begin{tabular}[x]{@{}c@{}} {$  \color[rgb]{0.8070588235294117,0.9490196078431372,0.19137254901960785} \Large  \nicefrac{\bm{3}}{\bm{5}} $}\\ {\color{gray} \tiny (3, 0, 0, 2)}\end{tabular} & \begin{tabular}[x]{@{}c@{}} {$  \color[rgb]{0.6784313725490196,0.45647058823529413,0.11764705882352941} \Large  \nicefrac{\bm{1}}{\bm{5}} $}\\ {\color{gray} \tiny (1, 0, 3, 1)}\end{tabular} & \begin{tabular}[x]{@{}c@{}} {$  \color[rgb]{0.4823529411764706,0.09411764705882353,0.09411764705882353} \Large  \nicefrac{\bm{0}}{\bm{5}} $}\\ {\color{gray} \tiny (0, 0, 4, 1)}\end{tabular} & \begin{tabular}[x]{@{}c@{}} {$  \color[rgb]{0.4823529411764706,0.09411764705882353,0.09411764705882353} \Large  \nicefrac{\bm{0}}{\bm{5}} $}\\ {\color{gray} \tiny (0, 0, 0, 5)}\end{tabular} & \begin{tabular}[x]{@{}c@{}} {$  \color[rgb]{0.4823529411764706,0.09411764705882353,0.09411764705882353} \Large  \nicefrac{\bm{0}}{\bm{5}} $}\\ {\color{gray} \tiny (0, 0, 0, 5)}\end{tabular}\\
{ Ideal Gas} & \begin{tabular}[x]{@{}c@{}} {$  \color[rgb]{0.1450980392156863,0.7450980392156863,0.34509803921568627} \Large  \nicefrac{\bm{5}}{\bm{5}} $}\\ {\color{gray} \tiny (5, 0, 0, 0)}\end{tabular} & \begin{tabular}[x]{@{}c@{}} {$  \color[rgb]{0.4823529411764706,0.09411764705882353,0.09411764705882353} \Large  \nicefrac{\bm{0}}{\bm{5}} $}\\ {\color{gray} \tiny (0, 0, 0, 5)}\end{tabular} & \begin{tabular}[x]{@{}c@{}} {$  \color[rgb]{0.1450980392156863,0.7450980392156863,0.34509803921568627} \Large  \nicefrac{\bm{5}}{\bm{5}} $}\\ {\color{gray} \tiny (5, 0, 0, 0)}\end{tabular} & \begin{tabular}[x]{@{}c@{}} {$  \color[rgb]{0.4823529411764706,0.09411764705882353,0.09411764705882353} \Large  \nicefrac{\bm{0}}{\bm{5}} $}\\ {\color{gray} \tiny (0, 0, 4, 1)}\end{tabular} & \begin{tabular}[x]{@{}c@{}} {$  \color[rgb]{0.4823529411764706,0.09411764705882353,0.09411764705882353} \Large  \nicefrac{\bm{0}}{\bm{5}} $}\\ {\color{gray} \tiny (0, 0, 0, 5)}\end{tabular} & \begin{tabular}[x]{@{}c@{}} {$  \color[rgb]{0.4823529411764706,0.09411764705882353,0.09411764705882353} \Large  \nicefrac{\bm{0}}{\bm{5}} $}\\ {\color{gray} \tiny (0, 0, 0, 5)}\end{tabular}\\
{ Rydberg} & \begin{tabular}[x]{@{}c@{}} {$  \color[rgb]{0.4823529411764706,0.09411764705882353,0.09411764705882353} \Large  \nicefrac{\bm{0}}{\bm{5}} $}\\ {\color{gray} \tiny (0, 0, 0, 5)}\end{tabular} & \begin{tabular}[x]{@{}c@{}} {$  \color[rgb]{0.4823529411764706,0.09411764705882353,0.09411764705882353} \Large  \nicefrac{\bm{0}}{\bm{5}} $}\\ {\color{gray} \tiny (0, 0, 0, 5)}\end{tabular} & \begin{tabular}[x]{@{}c@{}} {$  \color[rgb]{0.4823529411764706,0.09411764705882353,0.09411764705882353} \Large  \nicefrac{\bm{0}}{\bm{5}} $}\\ {\color{gray} \tiny (0, 0, 5, 0)}\end{tabular} & \begin{tabular}[x]{@{}c@{}} {$  \color[rgb]{0.4823529411764706,0.09411764705882353,0.09411764705882353} \Large  \nicefrac{\bm{0}}{\bm{5}} $}\\ {\color{gray} \tiny (0, 0, 0, 5)}\end{tabular} & \begin{tabular}[x]{@{}c@{}} {$  \color[rgb]{0.4823529411764706,0.09411764705882353,0.09411764705882353} \Large  \nicefrac{\bm{0}}{\bm{5}} $}\\ {\color{gray} \tiny (0, 0, 0, 5)}\end{tabular} & \begin{tabular}[x]{@{}c@{}} {$  \color[rgb]{0.4823529411764706,0.09411764705882353,0.09411764705882353} \Large  \nicefrac{\bm{0}}{\bm{5}} $}\\ {\color{gray} \tiny (0, 0, 0, 5)}\end{tabular}\\
\bottomrule\end{tabularx}
\unskip\label{output/results.tex}\unskip%

    \caption{Results of each algorithm on \bench.
    The fraction given is the number of correct expressions rediscovered, divided by the number of total trials.
    In parentheses, a detailed itemization of the five trials is given, in the order: 1) the number of correct rediscoveries, 2) the number of nearly-correct rediscoveries, 3) the number of runs which failed to produce a well-defined expression, and 4) the number of incorrect rediscoveries.
    }
    \label{tab:results}
\end{table*}

For each of the 9 equations in \bench, we run 5 trials for each of the 6 algorithms, and record the Pareto front of each trial.
Each algorithm is fed the entire dataset for training.
This is because our \textit{test is the output expression}, rather than a separate test dataset.
Following this, each of the Pareto fronts was analyzed, by eye, to see whether the true expression was contained within it.
This was performed manually, as oftentimes automated equality checking with \texttt{sympy} produced incorrect results.
Furthermore, checking expressions by eye allows for small errors in recovered expressions to be ignored, so long as the overall functional form is correct.
It is also superior to numerically checking expressions, as overcomplicated but accurate expressions can be correctly penalized.

For this analysis, each Pareto front was assigned to one of four categories: (1) \textit{correct} if the Pareto front contained the true expression, (2) \textit{almost} if one of the expressions in the Pareto front was off of the true expression by a constant factor somewhere, (3) \textit{failed}, if the search produced no results (e.g., due to numerical instability, segmentation fault, domain errors, etc.), and (4) \textit{incorrect} if the true expression was not found in the Pareto front.
All raw data with all discovered Pareto fronts for this analysis is available online at \myhref{https://github.com/MilesCranmer/pysr_paper}{github.com/MilesCranmer/pysr\_paper}.

We emphasize that the stability of an SR algorithm and software implementation itself has central importance to practical use in science, and therefore, unstable runs which either produce an undefined expression, or crash, are \textit{included} in the results dataset and the primary score.
However, one should also note that because we are the authors of \pysr, we are therefore more likely to be running in a stable environment than, say, a package which we are unfamiliar with.
For example, as seen in \cref{tab:results}, DSR sees improved results if one ignores the failed runs, although there were still some expressions it struggled on.
Thus, detailed itemization of these results is given in \cref{tab:results}, to allow for a more nuanced understanding of the results.
Also note that this benchmark does not explicitly measure other metrics which are considered in other benchmarks, such as number of evaluations (where, e.g., genetic algorithms are very inefficient), nor does it consider the numerical accuracy of the expressions.

\subsection{Discussion}

It is also very important to note that these results are from the datasets specific to this competition.
Many of these algorithms may do very well against synthetic datasets generated for each expression.
For example, Hubble's law is simply a linear relation, so it could be surprising that many algorithms are not able to find it.
However, to perform well at the benchmark, each algorithm must find these expressions from the noisy and sometimes biased datasets shown in \cref{fig:bench_data}.

What is also interesting is that the two pure deep learning approaches, EQL and SR-Transformer, recovered the fewest expressions out of all tested algorithms.
One of these, EQL, learns the expression in an online fashion, while SR-Transformer performs fast inference based on pre-trained weights.
SR-Transformer, in particular, is pre-trained on billions of randomly-generated synthetic expressions on a cluster of GPUs for weeks.
However, the ``untrained'' algorithms written using classic heuristics: PySR, Operon, DSR, and QLattice, all out-performed these modern deep learning models. 
Perhaps this is an insight into the difficulty of preparing deep learning systems for real-world data and unexpected ``long-tail'' examples, whereas handwritten standard algorithms will often perform just as well on unexpected examples.  In SR especially, the space of potential expressions is so massive and nonlinear that interpolation techniques might suffer; consider that $x/y$ and $x\times y$ are just a single mutation away, but infinitely distinct in their data representation.
In spite of this, while pure deep learning strategies appears to require improvements in architecture or training to be useful in a real world SR setting, its success in, e.g.,~\cite{cranmerDiscoveringSymbolicModels2020,petersenDeepSymbolicRegression2021,kamiennyEndtoendSymbolicRegression2022} shows there is indeed strong potential long-term for such hybrid methods, and these directions should continue to be pursued.

\section{Conclusion}

In this paper, we have presented \pysr, an open-source library for practical symbolic regression.
We have developed \pysr with the aim of democratizing and popularizing symbolic regression in the sciences.
By leveraging a powerful multi-population evolutionary algorithm, a unique evolve-simplify-optimize loop, a high-performance distributed backend, and integration with deep learning packages, \pysr is capable of accelerating the discovery of interpretable symbolic models from data.

Furthermore, through the introduction of a new benchmark, \bench, we have provided a means to quantitatively evaluate the performance of symbolic regression algorithms in scientific applications.

Our results indicate that, despite advances in deep learning, classic algorithms build on evolution and other untrained symbolic techniques such as PySR, Operon, DSR, and QLattice, still outperform pure deep-learning-based approaches, EQL and SR-Transformer, in discovering historical empirical equations from original and synthetic datasets.
This highlights the challenges faced by deep learning methods when applied to real-world data with its biases and heteroscedasticity, and the nonlinearity of the space of expressions.

However, we emphasize that deep learning techniques---inclusive of generative models like~\cite{kamiennyEndtoendSymbolicRegression2022}, reinforcement learning methods such as~\cite{petersenDeepSymbolicRegression2021}, and symbolic distillation~\cite{cranmerDiscoveringSymbolicModels2020}---all hold potential for improving symbolic regression, and encourage further exploration of such hybrid methods.

Over the past several years since the release of \pysr in 2020, there have been a number of exciting \pysr applications to discover new models in various subfields. There are too many to list here in detail, but we list some examples:~\cite{grundnerDataDrivenEquationDiscovery2023} use \pysr to discover a new symbolic parameterization for cloud cover formation;~\cite{liElectronTransferRules2023} use \pysr alongside other ML techniques to discover electron transfer rules in various materials;~\cite{matchevAnalyticalModelingExoplanet2022,wadekarSZFluxmassYM2022,wadekarAugmentingAstrophysicalScaling2022} combine \pysr with dimensional analysis to discover new astrophysical relations;~\cite{verstyukMachineLearningGravity2022} use \pysr in economics, to find effective rules governing international trade;~\cite{kidgerNeuralDifferentialEquations2022} demonstrates how to use \pysr to extract dynamical equations learned by a neural differential equation; and, finally,~\cite{wongAutomatedDiscoveryInterpretable2022} use \pysr to discover interpretable population models of gravitational wave sources.

In some ways this list of recent applications provides the strongest validation yet of the science and user-focused approach we argued for in \cref{sec:intro_pysr}, as \pysr has already been applied successfully to model discovery in a variety of fields.
It is our hope that \pysr will continue to grow as a community tool, and provide value to researchers, helping discover interpretable symbolic relationships in data and ultimately leading to new insights, theories, and advancements in their respective fields.

\paragraph{Acknowledgements}
This software was built in the Python~\cite{vanrossumPythonReferenceManual2009} and Julia~\cite{julia} programming languages.
Direct dependencies of PySR include
\texttt{numpy}~\cite{numpy},
\texttt{sympy}~\cite{sympy},
\texttt{sklearn}~\cite{sklearn},
and
\texttt{pandas}~\cite{pandas},
with export functionality provided by
\texttt{jax}~\cite{jax}
and
\texttt{pytorch}~\cite{torch}.
Key dependencies of SymbolicRegression.jl include
\texttt{Optim.jl}~\cite{mogensenOptimMathematicalOptimization2018}, \texttt{LoopVectorization.jl}~\cite{elrodLoopVectorizationJlMacro2022}, \texttt{Zygote.jl}~\cite{innesDonUnrollAdjoint2018}, and
\texttt{SymbolicUtils.jl}~\cite{gowdaHighperformanceSymbolicnumericsMultiple2022}.
The packages \texttt{matplotlib}~\cite{matplotlib} and
\texttt{showyourwork}~\cite{lugerShowyourwork2021}
were also used in producing this manuscript.
Quanta magazine~\cite{woodPowerfulMachineScientists2022} was used as artistic inspiration for some figures.

Miles Cranmer would like to thank
the Simons Foundation for providing resources for pursuing this research;
Shirley Ho and David Spergel for countless insightful discussions about \pysr, feedback on this manuscript, promotion of it as a tool in the sciences, and for their support of this project;
my research collaborators who provided feedback throughout the development of \pysr, including Pablo Lemos, Peter Battaglia, Steve Brunton, Jay Wadekar, Paco Villaescusa-Navarro, Kaze Wong, Elaine Cui, Christina Kreisch, Nathan Kutz, Drummond Fielding, Keaton Burns, Dima Kochkov, Alvaro Sanchez-Gonzalez, Christian Jespersen, Patrick Kidger, Kyle Cranmer, Niall Jeffrey, Ana Maria Delgado, Keming Zhang, Pierre-Alexandre Kamienny, Michael Douglas, Francois Charton;
all the wonderful open-source code contributors, including Mark Kittisopikul, T Coxon, Dhananjay Ashok, Johan Blåbäck, Julius Martensen, GitHub user \texttt{@ngam}, Christopher Rackauckas, Jerry Ling, Charles Fox, Johann Brehmer, Marius Millea, GitHub user \texttt{@Coba}, Pietro Monticone, Mateusz Kubica, GitHub user \texttt{@Jgmedina95}, Michael Abbott, Oscar Smith, and several others;
Marco Virgolin for extremely helpful comments on a draft of this paper, as well as general feedback;
Bill La Cava for providing feedback as well as spearheading the SRBench initiative, along with the rest of the SRBench organizers;
Brenden Petersen for feedback on \pysr as well as providing insights discussions about the SR landscape;
and so many others who have provided support to the project through email, Twitter, GitHub issues, and in-person.
I am blown away by the community that is forming around \pysr and the positive feedback it has received. Thank you.

\bibliographystyle{unsrt85}
\bibliography{ms}

\end{document}